\documentclass{aa}  

\usepackage{graphicx}
\usepackage[varg]{txfonts}

\usepackage{lscape}

\defcitealias{clem}{CLHW11}
\begin{document}

   \title{Spectroscopic membership for the populous 300\,Myr-old open cluster NGC~3532
		 	\thanks{Based on data acquired through the Australian Astronomical Observatory, under program S/2017A/02.}
		 	\thanks{Based on observations at Cerro Tololo Inter-American Observatory, National Optical Astronomy Observatory under proposals 2008A-0476, 2008A-0512, 2008B-0248, 2010A-0281, 2010B-0492, and 2011B-0322.}
		 	\thanks{The full Tables 1, 3, and 4 are only available in electronic form
		 		at the CDS via anonymous ftp to cdsarc.u-strasbg.fr (130.79.128.5)
		 		or via http://cdsweb.u-strasbg.fr/cgi-bin/qcat?J/A+A/}
		 } 
 	
   \author{D. J. Fritzewski\inst{1}
          \and
          S. A. Barnes\inst{1,2}
          \and
          D. J. James\inst{3}
          \and
          A. M. Geller\inst{4,5}
          \and
          S. Meibom\inst{6}
          \and
          K. G. Strassmeier\inst{1}
          }

   \institute{Leibniz-Institut f\"ur Astrophysik Potsdam (AIP),
              An der Sternwarte 16, 14482 Potsdam, Germany\\
              \email{dfritzewski@aip.de}
         \and
	         Space Science Institute, 4750 Walnut St., Boulder, CO 80301, USA
         \and
             Event Horizon Telescope, Harvard-Smithsonian Center for Astrophysics, 60 Garden Street, Cambridge, MA 02138, USA
         \and
	         Center for Interdisciplinary Exploration and Research in Astrophysics (CIERA) and Department of Physics and Astronomy, Northwestern University, 2145 Sheridan Rd, Evanston, IL 60208, USA
	     \and
		     Adler Planetarium, Department of Astronomy, 1300 S. Lake Shore Drive, Chicago, IL 60605, USA
		 \and
			 Harvard-Smithsonian Center for Astrophysics, 60 Garden Street, Cambridge, MA 02138, USA
             }

   \date{}

 
  \abstract
   {NGC~3532 is an extremely rich open cluster embedded in the Galactic disc, hitherto lacking a comprehensive, documented membership list.}
   {We provide membership probabilities from new radial velocity observations of solar-type and low-mass stars in NGC~3532, in part as a prelude to a subsequent study of stellar rotation in the cluster. }
   {Using extant optical and infra-red photometry we constructed a preliminary photometric membership catalogue, consisting of 2230 dwarf and turn-off stars.
   	We selected 1060 of these for observation with the AAOmega spectrograph at the 3.9\,m-Anglo-Australian Telescope and 391 stars for observations with the Hydra-South spectrograph at the 4\,m Victor Blanco Telescope, obtaining spectroscopic observations over a decade for 145 stars.
   	We measured radial velocities for our targets through cross-correlation with model spectra and standard stars, and supplemented them with radial velocities for 433 additional stars from the literature.
    We also measured $\log g$, $T_\mathrm{eff}$, and [Fe/H] from the AAOmega spectra.}
    {The radial velocity distribution emerging from the observations is centred at $5.43\pm0.04$\,km\,s$^{-1}$ and has a width (standard deviation) of 1.46\,km\,s$^{-1}$.
    Together with proper motions from \emph{Gaia} DR2 we find 660 exclusive members, of which five are likely binary members.
    The members are distributed across the whole cluster sequence, from giant stars to M dwarfs, making NGC~3532 one of the richest Galactic open clusters known to date, on par with the Pleiades.
    From further spectroscopic analysis of 153 dwarf members we find the metallicity to be marginally sub-solar, with $\mathrm{[Fe/H]}=-0.07\pm0.10$.
    We confirm the extremely low reddening of the cluster, $E_{B-V} = 0.034\pm0.012$\,mag, despite its location near the Galactic plane.
	Exploiting trigonometric parallax measurements from \emph{Gaia} DR2 we find a distance of $484^{+35}_{-30}$\,pc [$(m-M)_0=8.42\pm0.14$\,mag].
	Based on the membership we provide an empirical cluster sequence in multiple photometric passbands.
    A comparison of the photometry of the measured cluster members with several recent model isochrones enables us to confirm the 300\,Myr cluster age. However, all of the models evince departures from the cluster sequence in particular regions, especially in the lower mass range.}
   {}

   \keywords{Open clusters and associations: individual: NGC 3532 - Stars: late-type - Stars: abundances - Stars: kinematics and dynamics - Galaxy: Stellar content - Techniques: radial velocities}

   \titlerunning{Spectroscopic membership for NGC~3532}
   \authorrunning{D. J. Fritzewski et al.}

   \maketitle
%
\section{Introduction}


In the first recorded description \cite{LaCaille} called the southern open cluster \object{NGC~3532} a ``prodigious cluster of small stars''\footnote{``Amas prodigieux de petites \'etoiles [\dots]''} and for \cite{herschel1847} NGC~3532 was ``the most brilliant object of the kind I have ever seen''. However, NGC~3532 has received little attention in modern times compared with other open clusters in the southern sky. Our radial velocity study confirms Herschel's statement, and this paper is the first in a series that we plan, aimed at transforming NGC~3532 into a canonical young open cluster that can be used by the community for new and detailed astrophysical investigations, e.g. on the rotational dynamo transitions of cool stars (e.g. \citealt{2003ApJ...586..464B}).

As regards prior work, an initial photoelectric photometry study was published by \cite{1959BAN....14..265K}, and despite the limited magnitude range of the sample the author was able to derive a distance of 432\,pc. \cite{1980A&AS...39...11F} presented additional photoelectric and photographic photometry for 700 stars in the region of NGC~3532. This was soon followed by a radial velocity study including the Koelbloed sample published by \cite{1980A&AS...41..245G, 1981A&A....99..155G}. Further photometric studies in different filters and with additional targets included \cite{1981A&AS...43..421J}, \cite{1981ApJ...246..817E}, \cite{1982AJ.....87.1390W}, \cite{1987A&AS...71..147S}, and \cite{1988MNRAS.235.1129C}. Those studies concentrated on the upper main sequence and the giant branch stars. Surprisingly, no further photometric study of main sequence stars in NGC~3532 was published for 23 years\footnote{The otherwise unpublished PhD thesis of one of the authors \citep{1997PhDT.........7B} includes standardised and time-series photometry of NGC~3532.} until the recent comprehensive CCD-photometric study of \cite{clem}, hereafter referred to as C11. They measured the magnitudes and positions of stars to a depth of $V=22$ in a 1\degr{} field of view.

The distance to NGC~3532 has been estimated using photometry, and also has been directly measured by the Hipparcos mission. The initial reduction, based on eight stars, \cite{1999A&A...345..471R}, yielded  $405^{+76}_{-55}$\,pc. The precision was improved in the new Hipparcos catalogue \citep{2009A&A...497..209V}, and found to be essentially the same, $406^{+75}_{-56}$\,pc, based on six stars. In contrast isochrone fitting to the cluster sequence gave a distance of $492^{+12}_{-11}$\,pc (C11). From parallax measurements included in the second \emph{Gaia} data release (\emph{Gaia} DR2, \citealt{2018A&A...616A...1G}), \cite{2018A&A...616A..10G} found 484\,pc, reducing the tension considerably.

The age of the open cluster has been estimated by most of the photometric studies. \cite{1959BAN....14..265K} suggested 100\,Myr, based on the absolute magnitude-age relation by \cite{1957ApJ...125..435S}. Both \cite{1980A&AS...39...11F} and \cite{1981A&AS...43..421J} estimated 200\,Myr, while \cite{1981ApJ...246..817E} gave 350\,Myr. More recent estimates based on isochrone fitting are 300\,Myr from C11 and 310\,Myr by \cite{2012A&A...541A..41M}. Based on white dwarfs in the cluster and mostly independent of the main sequence stars, \cite{2012MNRAS.423.2815D} constrained the age to $300\pm25$\,Myr. Modern age estimates for NGC~3532 are all in agreement with 300\,Myr and our work will be shown to confirm this.

With the notable exception of the above-mentioned studies by \cite{1980A&AS...41..245G, 1981A&A....99..155G}, who obtained objective prism spectra of 84 main sequence stars near the turn-off, most radial velocity studies of the open cluster are from recent times. \cite{2002AJ....123.3318G} observed 21 of the brightest cluster members and \cite{2008A&A...485..303M} focused on the cluster giants (8 stars). The latest public spectroscopic surveys RAVE \citep{2006AJ....132.1645S}, \emph{Gaia}-ESO \citep{2012Msngr.147...25G}, and \emph{Gaia} \citep{2016A&A...595A...1G} included some cluster stars but did not specifically investigate the cluster sequence. 

The richness of NGC~3532 may also be observed in the number of known cluster white dwarfs. \cite{1989A&A...218..118R} discovered seven white dwarf candidates on photographic plates of which three could be verified spectroscopically \citep{1993A&A...275..479K}. More recently \cite{2009MNRAS.395.2248D, 2012MNRAS.423.2815D} have confirmed further white dwarfs and \cite{2016MNRAS.457.1988R} have discovered an additional candidate. Having seven confirmed white dwarf members makes NGC~3532 a rich source for degenerate objects among Galactic open clusters \citep{2012MNRAS.423.2815D}, further testifying to its value for studies of stellar evolution.

The overall richness of the cluster is apparent in the photometric work by  \cite{1987A&AS...71..147S} and C11, showing NGC~3532 to have a cluster sequence extending deep enough for it to be embedded in the Galactic background field. We aim to identify the cluster members down to its M~dwarfs and out to 1\degr{} (diameter), in order to construct the corresponding clean cluster sequence.

Finally, we note that with an ecliptic latitude of $-56.4\degr{}$ (J2000), NGC~3532 will be in the 54-day observing window of NASA's Transiting Exoplanet Survey Satellite (TESS) mission \citep{2015JATIS...1a4003R}. In the era of exoplanet discovery and characterization, exoplanet hosts in open clusters have particularly high scientific worth because their ages and other properties can be relatively well-determined and ranked (e.g. \citealt{1958RA......5...41S}, \citealt{1964ApJ...140..544D}, \citealt{1993A&AS...98..477M}, \citealt{1996ApJ...458..600B}, \citealt{1996ApJ...469L..53R}, \citealt{2003ApJ...586..464B, 2007ApJ...669.1167B}). Knowing the ages of exoplanets and their host stars is of course crucial to better understand their physics and formation history.

This paper is structured as follows. In Sec.~\ref{sec:photmem} we construct and present the photometric membership of NGC~3532, enabling the selection of appropriate targets for spectroscopic study. In Sec.~\ref{sec:RV} and Sec.~\ref{sec:stellpar} we explain our data reduction, leading to the measurement of radial velocities and stellar parameters, respectively. In Sec.~\ref{sec:clseq} we present the cluster membership, the cluster sequence it defines, and the comparison with isochrones.

\section{Joint optical and infrared photometric membership}
\label{sec:photmem}

\begin{figure}
	\includegraphics[width=\columnwidth]{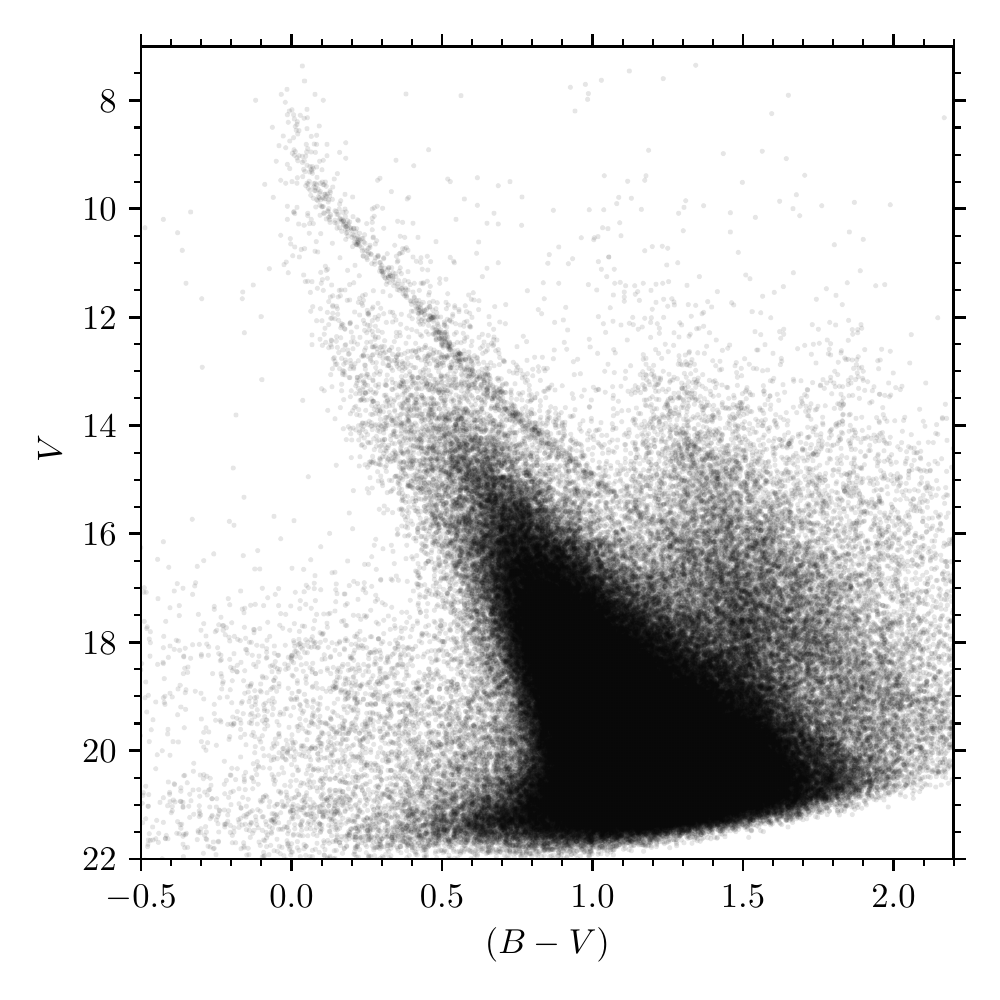}
	\caption{The (B, V) colour-magnitude diagram of the NGC~3532 field, based on photometry from C11. A rich and clear cluster sequence is visible against the substantial Galactic disc background field, the latter overpowering it below $V \sim 16$.}
\label{fig:simpCMD}
\end{figure}

The $(B,V)$ colour-magnitude diagram (CMD) in Fig.~\ref{fig:simpCMD} shows the full stellar content of the probed region, heavily contaminated with background stars. Because a membership list for NGC~3532 is not available in the literature, we constructed a list of joint optical and infrared photometric members. Although we used the photometry of C11 and followed their method for the construction of the single-star photometric cluster sequence, we were careful to also retain the potential photometric binary members.

C11 plotted colour-colour diagrams in [$(V-I_c)$, $(V-J)$] and [$(V-I_c)$, $(V-K_s)$], allowing the foreground dwarfs and the background giants to be separated based on the relations of \cite{1988PASP..100.1134B} and the appropriate reddening vector. Stars more than 0.5\,mag away in colour from the visible cluster sequence were rejected, and after transforming the distance of each star from the sequence into a $\chi^2$-value, they suppressed stars with $\chi^2$ greater than a threshold, and plotted the results in their paper. However, they did not provide the resulting list of candidate members.

\begin{figure}
	\includegraphics[width=\columnwidth]{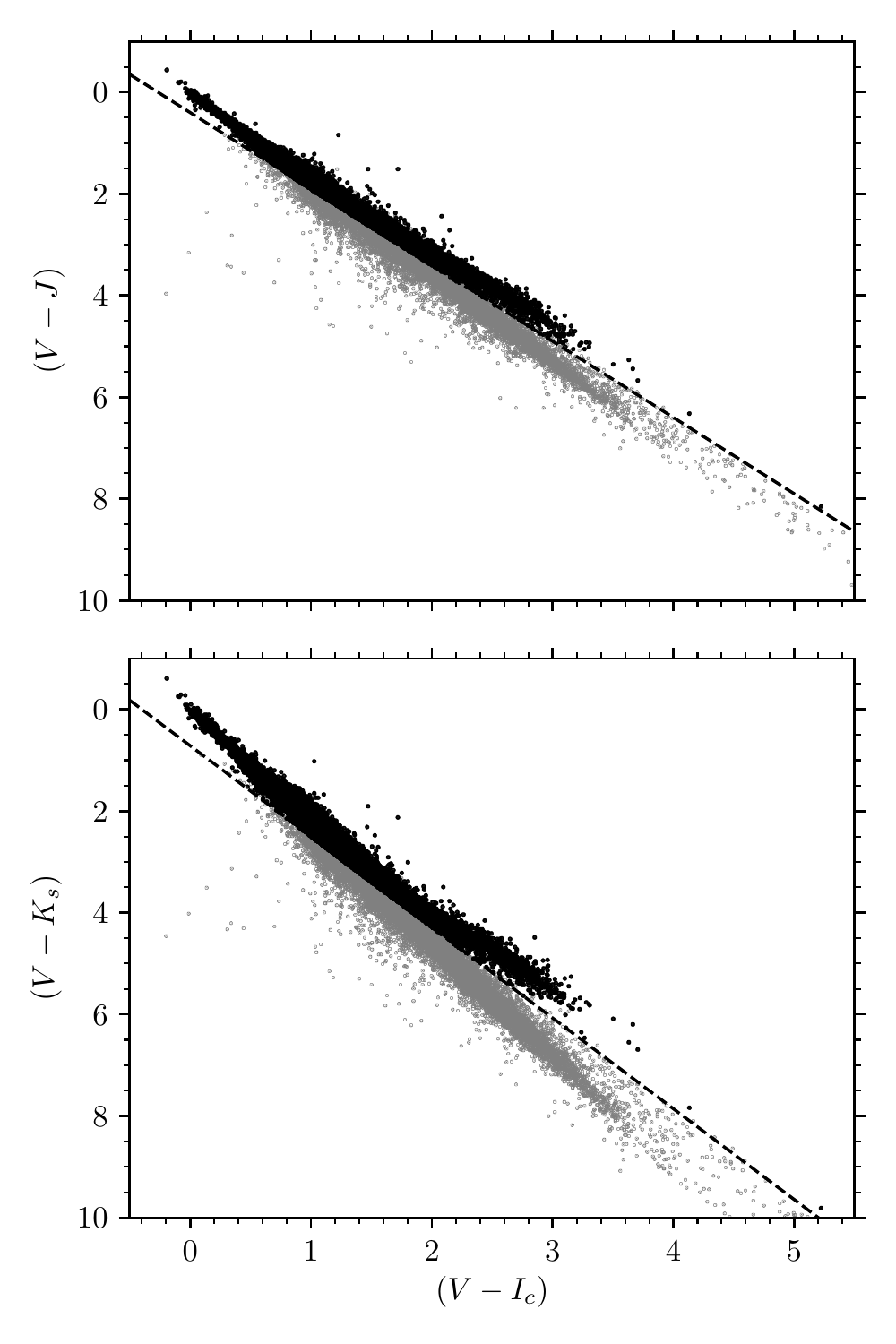}
	\caption{Colour-colour diagrams for the NGC~3532 field, showing the lines (see text) separating the retained (above) and suppressed (below) stars. The stars below the indicated lines, are reddened significantly more than the cluster stars and are presumed to be field stars.}
	\label{fig:colcol}
\end{figure}

We followed the general method of C11, but made certain deviations to suit our purposes, particularly to retain any cluster photometric binaries. We began by matching the $BV(RI)_c$ photometry from C11 with $JHK_s$ photometry from 2MASS \citep{2006AJ....131.1163S}, to construct colour-colour plots in [$(V-I_c)$, $(V-J)$] and [$(V-I_c)$, $(V-K_s)$] for all stars common to the two catalogues (see Fig.~\ref{fig:colcol}). While all stars from C11 were used, we included only those stars from the 2MASS photometry that have a photometric quality flag of A, B, C, or D for all three passbands. With these requirements we retained only the stars with valid detections in all photometric bands. This usage of 2MASS photometry imposes a new brightness limit on our membership, fortunately not adversely impacting our own scientific goals. The cluster sequence in Johnson-Cousins colours can be traced down to at least $V=21$ but 2MASS is limited to $J=15.8$, corresponding to $V\approx19$ (based on the YaPSI isochrones, \citealt{2017ApJ...838..161S}) for a cluster of NGC~3532's estimated distance and age (492\,pc, 300\,Myr, C11).

As suggested by C11, we included only stars above both dividing lines
\begin{equation}
 (V-J) = \frac{3}{2} (V-I_c) + \frac{2}{5} 
\end{equation}
and
\begin{equation}
(V-K_s) = \frac{25}{14}  (V-I_c) + \frac{5}{7},
\end{equation}
separating foreground and background stars (Fig.~\ref{fig:colcol}). This is possible for NGC~3532 because of the large difference in reddening between the cluster and the field. The cluster itself has low extinction, with $A_V = 0.087$\,mag (C11) while the line-of-sight extinction is $A_{V,\mathrm{LOS}} = 3.04$\,mag \citep{2011ApJ...737..103S}. Although, the two sequences of foreground and background stars overlap, most background disc stars can be removed from the sample with this method. Stars in the overlap region of the two sequences could still lie on the cluster sequence in the CMD (especially background giants).

\begin{figure*}
	\includegraphics[width=\textwidth]{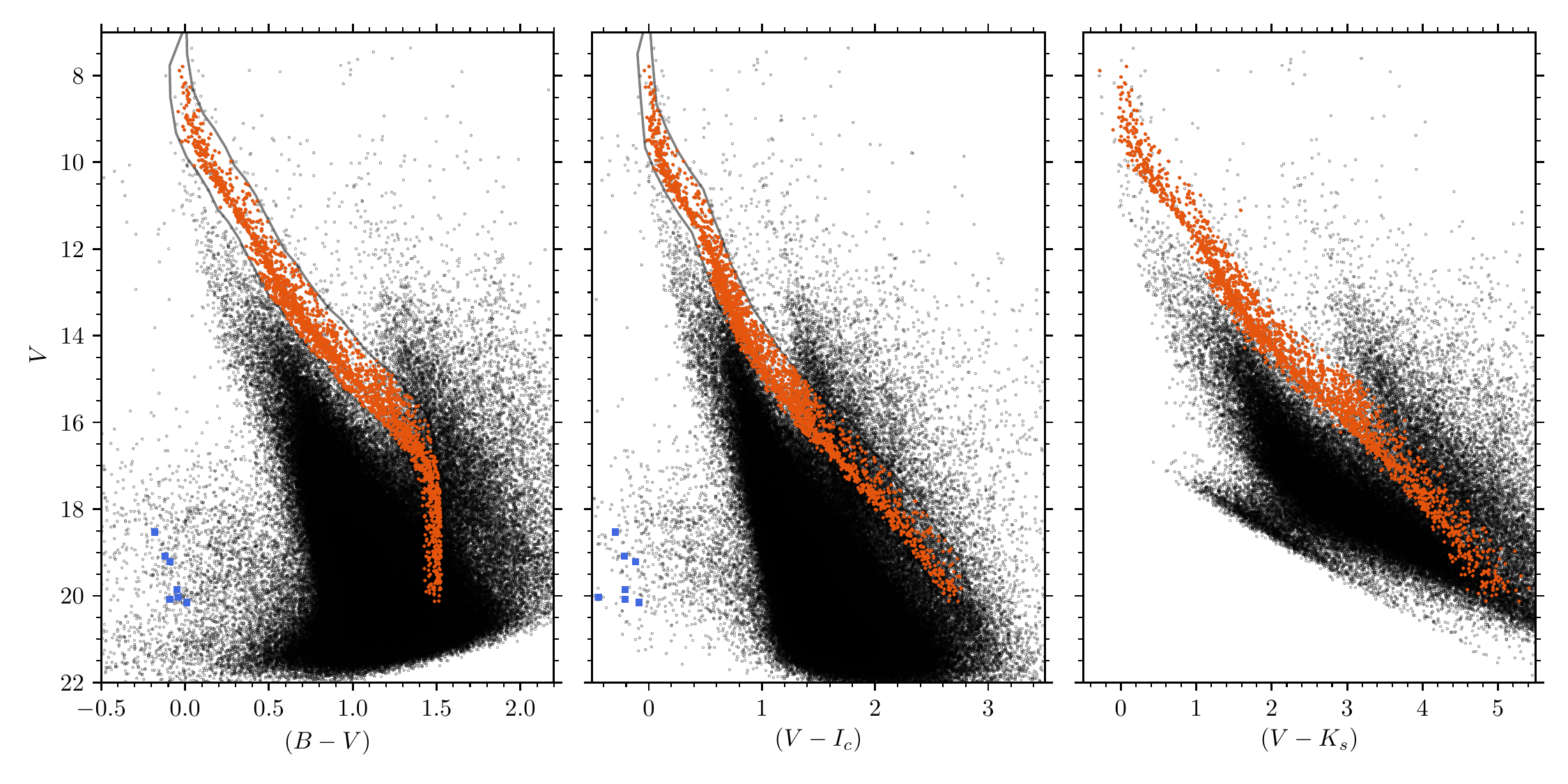}
	\caption{Colour-magnitude diagrams for NGC~3532. The left and centre panels include all stars in the C11 catalogue while the right panel shows a CMD in $(V-K_s)$ and $V$ with only the stars common to the 2MASS survey and the C11 sample. The grey stroke outlines the area defined as 0.1\,mag bluer and 1\,mag brighter than the hand-traced cluster sequence. In orange we show all stars that are both in the defined region, and are part of the dwarf sequence in Fig.~\ref{fig:colcol}. The seven confirmed white dwarfs in NGC~3532 are also marked (blue squares) in the first two panels.}
	\label{fig:cmd}
\end{figure*}

The cluster sequence is clearly visible in the cleaned CMD (discussed below), and can be traced by eye. In contrast to the exclusive selection of C11, we wish to retain the binary cluster members. Therefore, we do not calculate a $\chi^2$-value associated with the distance from the cluster sequence, and instead retain all stars up to 0.1\,mag bluer, and up to 1\,mag brighter than the hand-traced cluster sequence.
As such, our selection includes all potential members from the blue edge of the cluster sequence to the equal-mass binary sequence, the latter situated $\sim$0.75\,mag above the cluster sequence. These stars were extracted jointly from both the [$(B-V)$, $V$] and [$(V-I_c)$, $V$] CMDs (Fig.~\ref{fig:cmd}), and are hereafter called \textit{photometric cluster members}.
A [$(V-K_s)$, $V$] CMD (Fig.~\ref{fig:cmd}, right-most panel) was additionally used to verify that all photometrically-selected cluster members are indeed on the cluster sequence, including in colours not used to extract the members. We observe that this is indeed the case.
Within the 1\degr{} diameter limit for NGC~3532, we find a total of 2230 stars on the photometric cluster sequence common to both the C11 and 2MASS samples. These are listed in Table~\ref{tab:photomem}.

\begin{table*}
	\caption{Photometric dwarf members of NGC~3532 cross-matched with 2MASS and \emph{Gaia} DR2. The proper motion membership is from Sec.~\ref{sec:pm}.}
	\centering
	\label{tab:photomem}
	\begin{tabular}{llllllll}
	\hline
	\hline
	CLHW & RAJ2000 & DEJ2000 & \emph{Gaia} DR2 ID & 2MASS ID & $\mu_\alpha$ & $\mu_\delta$ & PM \\
	& ($\degr$) & ($\degr$) &  & & mas\,yr$^{-1}$ & mas\,yr$^{-1}$ & \\
	\hline
	85 & 167.42946 & -58.37269 & 5339436345251570944 & 11094305-5822217 & -10.3164 & 4.9711 & y\\
	206 & 167.42750 & -58.40275 & 5339430332297021568 & 11094258-5824099 & -5.0697 & 2.6553 & n\\
	451 & 167.41925 & -58.24475 & 5339438853512774656 & 11094060-5814411 & -10.7149 & 5.2366 & y\\
	485 & 167.42150 & -58.33967 & 5339436688849023232 & 11094115-5820228 & -4.0350 & 2.8414 & n\\
	530 & 167.42896 & -58.61464 & 5339403218113480192 & 11094293-5836527 & 2.9750 & -0.9000 & n\\
	\dots
	
\end{tabular}
\tablefoot{The full table is available at the CDS. \textit{CLHW}: ID from C11; proper motions from \emph{Gaia} DR2; \textit{PM}: proper motion member (y/n).}
\end{table*}

The next logical step would normally have been to refine the cluster sequence with the help of the proper motions of the stars on the preliminary (i.e. photometric) cluster sequence. As data from the \emph{Gaia} mission were not readily available at the time of target selection this did not yield any significant additional information because the mean proper motion of the cluster does not differ significantly from the background (c.f. C11 for a detailed analysis). Consequently, we did not include the proper motions and have solely used the photometric membership information to select the targets for the spectroscopic observations. Data from the \emph{Gaia} mission became available after our observing campaigns in the form of UCAC5 \citep{2017AJ....153..166Z} and \emph{Gaia} DR2 \citep{2018A&A...616A...1G}. We discuss the proper motion of the open cluster later, in Sec.~\ref{sec:pm}, in the light of those new data and our spectroscopic membership. We also note that initial \emph{Gaia}/TGAS proper motions \citep{2017A&A...601A..19G} appear to have found bright kinematic cluster members up to 5\degr{} from the cluster centre. Lacking dedicated cluster photometry for the outer regions, and with a focus on the inner parts of NGC~3532, we have not included those stars in our sample.

\section{Radial velocity measurements}
\label{sec:RV}

We targeted the photometric candidate members in spectroscopic observing campaigns at the Anglo-Australian Telescope and the Victor~Blanco Telescope. Here we present the observations, the resultant radial velocity distribution, compare our data to prior work, and include a relatively small number of additional measurements from the literature and public surveys.

\subsection{Observations}
\label{sec:obs}

\subsubsection{AAO}
NGC~3532 was observed on 10 and 11 March 2017 in service mode with the AAOmega two-armed spectrograph fed by the 392-fibre 2dF fibre positioner \citep{2002MNRAS.333..279L} at the 3.9\,m Anglo-Australian Telescope at Siding Spring observatory. We used the 1700D grating ($R=10\,000$) in the red arm, designed for radial velocity studies with the infra-red \ion{Ca}{ii} triplet (IRT) in the wavelength range from 8340\,\AA{} to 8940\,\AA. The blue arm, not useful to us because our targets are too faint for that spectral range, was equipped with the standard grating 580V ($R=1200$).

With a field of view of 2\degr{}, a large fraction of the cluster could be observed simultaneously, facilitating the target selection. We only targeted the photometric members of NGC~3532 as determined above, and assigned the highest priority to stars for which we have obtained differential photometry time-series\footnote{These data were obtained in order to study the rotation of the cluster's cool stars, and will be published separately (Fritzewski et al. in prep.).}. Those stars occupy a somewhat smaller field of view than the photometric study of C11. Correspondingly, lower observing priorities were assigned to stars that lie beyond our photometric monitoring region. However, the guide stars were assigned exclusively from the outer regions to avoid crowding in the inner part where our highest priority targets are situated.

The large field of view allowed us to divide the fibre configurations into two magnitude ranges to avoid fibre crosstalk. In total we observed NGC~3532 with three different fibre configurations, with the cluster centred in the field of view each time (equinox J2000.0, $\alpha=\mbox{11:05:39}$, $\delta=\mbox{-58:45:12}$, \citealt{2009MNRAS.399.2146W}). From the 392 fibres available 354 (352 in one case) fibres were used for the science targets, 25 were dedicated sky fibres, and 8 (7) fibres were used for the guiding stars. The remaining 13 (16) fibres were not used because they were either broken, or the field was too crowded to position them on a target.

The exposure time was 120\,min for the fainter stars ($I_c=15-17$, which corresponds to $V=16.6-19.5$) and 30\,min each for the two bright configurations ($I_c=12-15$, corresponding to $V=12.6-16.6$). Each exposure was split into three sub-exposures to assist in cosmic ray removal. For the bright configurations we were able to achieve signal-to-noise ratios (SNR) per pixel ranging from 20 for the faintest to 160 for the brightest stars. With the longer exposure time for the faint configuration we obtained SNRs between 10 and 80. Calibration files included dark, flat, bias, and arc frames. For the arc frames four lamps containing He, CuAr, FeAr, CuNe were used.

\begin{figure}
	\includegraphics[width=\columnwidth]{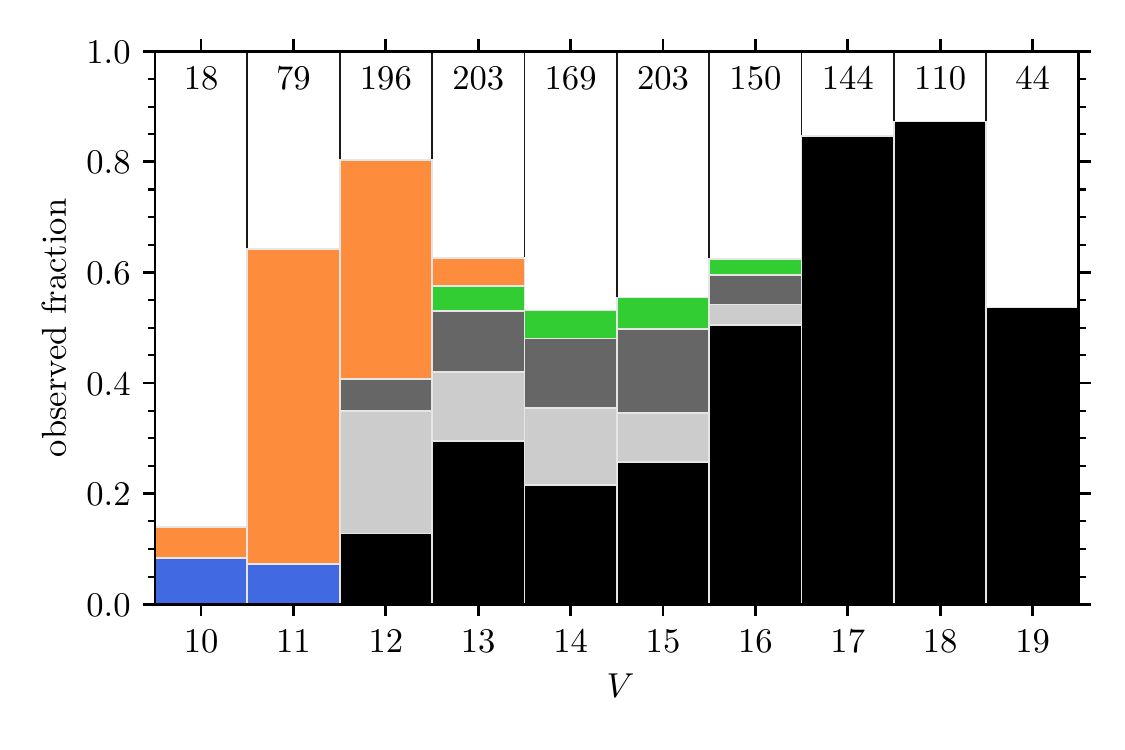}
	\caption{Stacked histogram of the fraction of observed photometric cluster members as a function of $V$. In blue we show the contribution from the RAVE survey (first two bins), in orange from \emph{Gaia} DR2 ($V=10$ to 13), and in green from the \emph{Gaia}-ESO survey ($V=12$ to 16). The components in black and light grey represent our own observations with only AAOmega and Hydra-S, respectively ($V=12$ to 19). In dark grey we show the contribution of stars observed with both instruments ($V=12$ to 16). The number in each above each bar indicate the total number of observed stars.}
	\label{fig:obs}
\end{figure}

With the time restrictions of service mode we limited our observations to three different fibre configurations. Hence, we were unable to observe all 2230 photometric members determined in Section~\ref{sec:photmem}, and observed only 1060 stars. The fraction of observed stars from the photometric cluster members is shown in Fig.~\ref{fig:obs} as a function of $V$. For most magnitude bins we were able to observe approximately half of the photometric members, sufficient for our purpose of constructing a cluster sequence.

\subsubsection{CTIO}
We also include previously-obtained (but hitherto unpublished) multi-epoch multi-object spectroscopy with the fibre-fed Hydra-S spectrograph at the Victor Blanco 4\,m Telescope at Cerro Tololo Interamerican Observatory (CTIO). The observations were carried out in seven runs over 15 nights from 28 February 2008 to 27 March 2010; an overview of these observations is given in Table~\ref{tab:Obslog}. We used Hydra-S in Echelle-mode with a wavelength-coverage from 5092\,\AA{} to 5274\,\AA{} ($\lambda_\mathrm{c}=5185\,\AA$, $R=18\,500$), a dispersion of 0.086\,\AA{}\,px$^{-1}$, and 1x2 binning. The observing strategy was similar to \cite{2008AJ....135.2264G} and we achieved a single measurement radial velocity precision of 0.8 km\,s$^{-1}$.

Fibres were placed on 391 stars selected photometrically from the colour-magnitude diagram in the range from $V=12$ to $V=16$. This sample has an overlap of 145 stars with the AAO sample. Each Hydra-S field was exposed for 2400\,s with up to four exposures in a sequence to remove cosmic rays from the data.

For calibration we obtained nightly ThAr arc frames, dark and bias frames, and flat fields in different configurations. The flat fields were both dome flats and sky flats. For the sky flats the fibres were either arranged in a circle or positioned as on target. We need these multi-configuration flat fields to remove systematics arising from the positioning of the fibres as detailed in \cite{2008AJ....135.2264G}. 

In total we obtained 1695 radial velocity measurements for 391 individual stars with Hydra-S. Multiple measurements of individual sources also allow us to identify binary stars from their radial velocity variability (see below).

\begin{table}
	\caption{Observing log for all observations at CTIO and AAO.}
	\label{tab:Obslog}
	\begin{tabular}{llll}
		\hline
		\hline
		Date & Observatory & Fields & Spectra\\
		\hline
		28 Feb. 2008 & CTIO & F1, F2 & 186\\ 
		22--24 Mar. 2008 & CTIO & F1--F4 & 337\\ 
		10/11 Jan. 2009 & CTIO & F1, F2 & 113\\
		30/31 Jan. 2009 & CTIO & F3, F4 & 185\\
		30/31 Jan. 2010 & CTIO & F1, F2 & 184\\
		27/28 Mar. 2010 & CTIO & F1--F4 & 335\\
		25--27 Jan. 2011 & CTIO & F1--F4& 355\\
		10/11 Mar. 2017 & AAO & 3 set-ups& 1060\\
		\hline
	\end{tabular}
\end{table}

\subsection{Data reduction and radial velocity determination}
\label{sec:rv}

To reduce and wavelength-calibrate the spectra obtained from AAOmega we used the standard 2dFdr pipeline (\citealt{2015ascl.soft05015A}, version 6.28). During pre-processing, the median bias and dark frames were subtracted from each science frame, and cosmic rays were removed. After fitting the image with a scattered light model the spectra were extracted. Those extracted spectra were divided by the fibre flat-fields and wavelength calibrated using $\sim$30 out of 40 available spectral lines in the arc frames. The sky lines were then used for throughput calibration and subsequently removed from the spectra. Finally, the three sub-exposures for each configuration were combined into a single spectrum.

Each combined spectrum was later continuum-normalized with a fifth-order Chebyshev polynomial. The radial velocity shift was measured with the \textsc{fxcor} routine from IRAF\footnote{IRAF is distributed by the National Optical Astronomy Observatories, which are operated by the Association of Universities for Research in Astronomy, Inc., under cooperative agreement with the National Science Foundation.} using the \textsc{PyRAF} interface. For the cross-correlation we used seven different template spectra from the synthetic spectral library of \cite{2014MNRAS.440.1027C}. For the hotter stars the infra-red triplet partly overlaps with lines from the Paschen series, which are not present for cooler stars. On the cool end of our observed sequence, the M dwarfs show TiO and VO bands. We assumed that our sample contains dwarf stars with solar-like metallicity and chose all template spectra to have $\log g = 4.5$ and $\mathrm{[Fe/H]} = 0$. To achieve a good match between the template spectrum and the observations we divided the data into magnitude bins and assigned a temperature from $T_\mathrm{eff}=$[7000, 6000, 5500, 5000, 4750, 4250, 3800]\,K to each of them. We have not used colour information because the effective temperature is monotonically dependent on the magnitude in the given range. The measured radial velocities were individually corrected for barycentric motion.

\begin{figure}
	\includegraphics[width=\columnwidth]{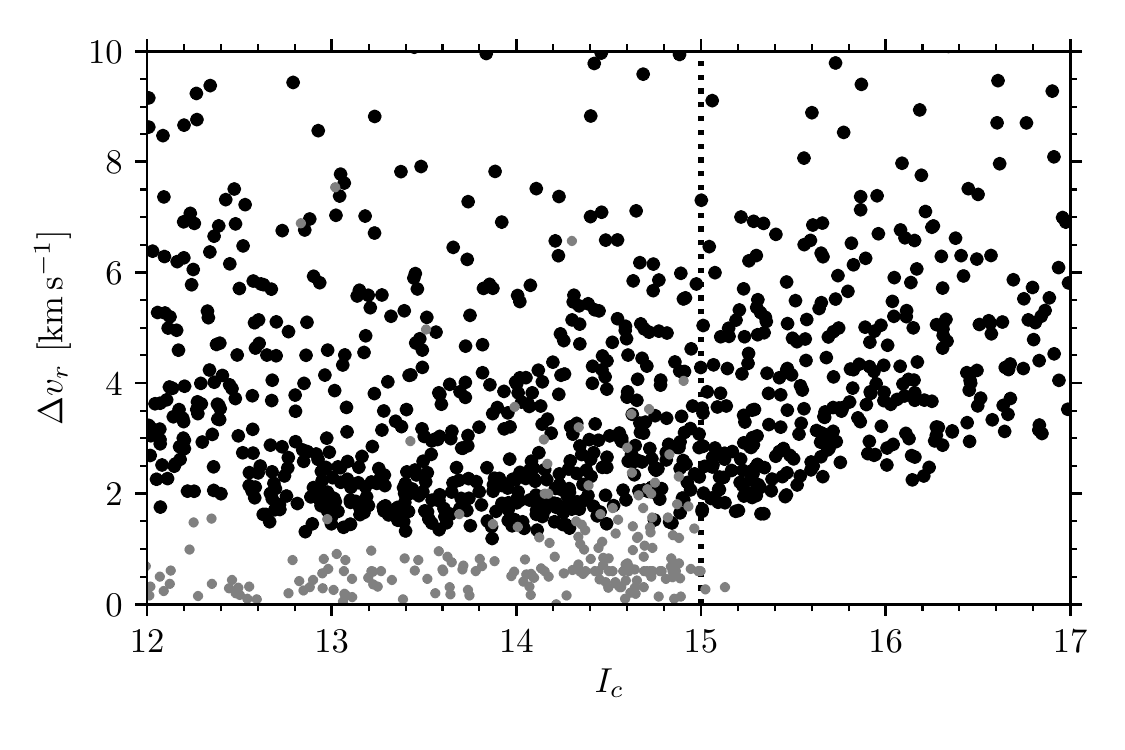}
	\caption{Distribution of the radial velocity uncertainties $\Delta v_r$ against the $I_c$ magnitude of the stars observed with AAOmega (black) and Hydra-S (grey). The dotted line shows the separation between the bright and faint configurations in the AAO campaign.}
	\label{fig:verror}
\end{figure}

The initial uncertainties obtained from IRAF were unrealistic. However, limiting the width of the Gaussian that IRAF uses to fit the cross-correlation peak to 10 pixels helps to achieve a good radial velocity measurement with a reasonable error estimate \citep{2016MNRAS.462.3376T}. The lower boundary of the error distribution, shown in Fig.~\ref{fig:verror}, is between 1.1\,km\,s$^{-1}$ and 3.2\,km\,s$^{-1}$ depending mostly on the spectral type; we show $I_c$ magnitude as a proxy. As seen from the smooth transition in Fig.~\ref{fig:verror} between the configurations with different exposure times (dotted line) the main contribution to the larger errors for the fainter stars can be attributed to the spectral type, rather than the SNR. Some stars show larger uncertainties because their cross-correlation functions are very wide and the heights are therefore lower. Hence, the height of the cross-correlation peak can be used as a quality indicator and larger radial velocity uncertainties are usually correlated with lower heights.

One reason for poor quality in certain cases could be fast rotation, and therefore broadened spectral lines. We also observe that for the hottest stars in our sample, e.g. those with very strong Paschen lines, the errors are much larger than for stars with weaker Paschen lines. In those cases no good match between the synthetic and the observed spectra could be achieved.

The median error for the (later determined) members of NGC~3532 with a good error estimate ($\Delta v_r < 10$\,km\,s$^{-1}$) is 2.0\,km\,s$^{-1}$ for the bright sample and 3.8\,km\,s$^{-1}$ for the fainter stars.

The Hydra-S observations were reduced and the radial velocities measured within IRAF. For details we refer the reader to the extensive description and error analysis in \cite{2008AJ....135.2264G}. The error distribution of these observations is constant for all magnitudes (see Fig.~\ref{fig:verror}) with a median uncertainty of 1.3\,km\,s$^{-1}$.

\subsection{Radial velocity distribution}
Before joining the two independently-reduced data sets we checked for systematic zero-point offsets. With the overlapping sample of 145 stars we are able to compare the zero-points directly. After removing all known radial velocity variables, 103 stars could be used to calculate the zero-point difference. We found no significant offset, hence we create a joint radial velocity distribution without adjustment.

In addition to our own two data sets we added the radial velocities from the \emph{Gaia}-ESO survey (GES), the RAVE survey DR5 \citep{2017AJ....153...75K} and \emph{Gaia} \citep{2018A&A...616A...5C}. The GES overlaps only with the AAO observations because the target region of the CTIO observations and the GES are different. We found only a minor zero-point offset ($\Delta v = 0.075\pm5.164$\,km\,s$^{-1}$) which is negligible for the subsequently determined membership probability (because the radial velocity distribution is much wider than the calculated offset). The RAVE data have no stars in common with any other data set because that survey targeted a different magnitude range, not permitting us to test for a zero-point offset. The radial velocities in the RAVE catalogue are similar to the data from all other observations (median 4.4\,km\,s$^{-1}$) and we can assume that any possible zero-point difference is small.

Although there is overlap between the \emph{Gaia} data and the other data sets, we have not used them to bring the data to a common level because the uncertainties in the Gaia data are much larger than any possible zero-point offset. For this reason we exclude the \emph{Gaia} radial velocities from the fit to the radial velocity distribution, while we include both of the other survey data sets from the surveys. Additionally, for stars with observations from multiple sources we chose ground-based data over \emph{Gaia} radial velocities.

\begin{figure}
	\includegraphics[width=\columnwidth]{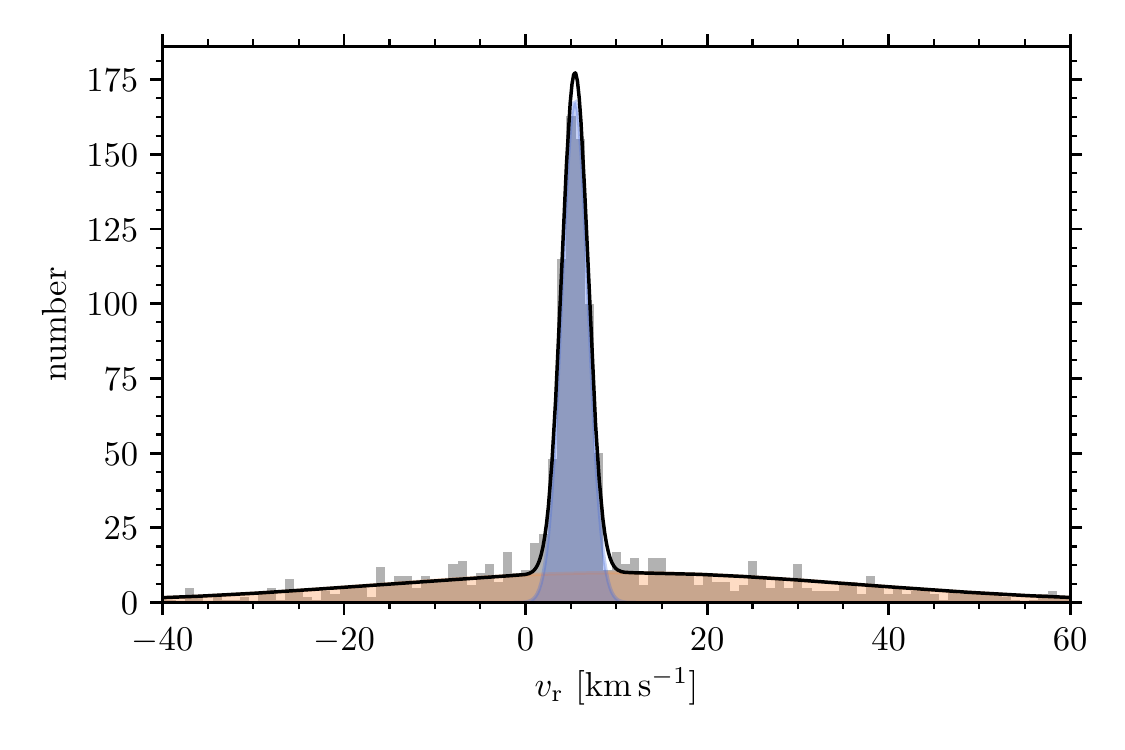}
	\caption{Radial velocity distribution of the combined data set. The fitted distribution is a two-component model with a mean cluster radial velocity of $v_\mathrm{r}=5.43$\,km\,$\mathrm{s}^{-1}$ ( $\sigma=1.46$\,km\,$\mathrm{s}^{-1}$), and a field component with $v_\mathrm{r}=10.3$\,km\,$\mathrm{s}^{-1}$ ( $\sigma=26$\,km\,$\mathrm{s}^{-1}$).}
	\label{fig:RVdistrib}
\end{figure}

Before fitting the radial velocity distribution with a model we excluded all known radial velocity variable stars (c.f. Sec.~\ref{sec:multiobs}) from our combined data set, noting that it could still include binary stars with only one data point. The cleaned sample should give the underlying cluster radial velocity distribution which we fit with a two-component Gaussian mixture model. It combines a strong cluster peak at $v_\mathrm{r}=5.43\pm{0.04}$\,km\,s$^{-1}$ ($\sigma_1=1.46$\,km\,s$^{-1}$) with a low-level, very wide ($v_\mathrm{r}=10.3\pm0.9$\,km\,s$^{-1}$, $\sigma_2=26$\,km\,s$^{-1}$) field component (Fig.~\ref{fig:RVdistrib}).

The width of the radial velocity distribution is a superposition of the intrinsic dispersion of the cluster, the width resulting from the measurement uncertainties, and some inflation from undetected binaries. Despite the combination of data with differing radial velocity precisions the width is dominated by the largest data set (AAOmega data, 59 per cent of the data points). Combining all data sets increases the width by less than five per cent.

\begin{figure}
	\includegraphics[width=\columnwidth]{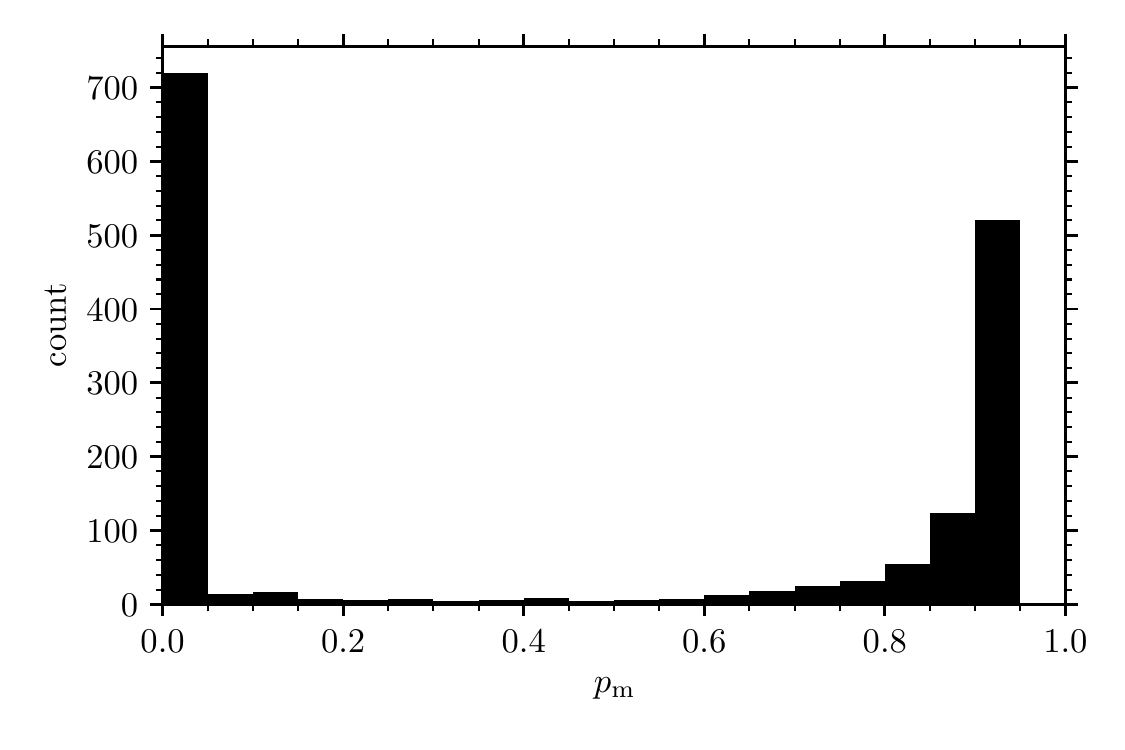}
	\caption{Membership probability distribution, $p_\mathrm{RV}$, obtained from the model in Fig.~\ref{fig:RVdistrib}. We include all stars with $p_\mathrm{RV} > 0.5$ as radial velocity members.}
	\label{fig:RVprob}
\end{figure}

With the obtained model ($\Phi(v_\mathrm{r})$) we calculate the membership probabilities $p_\mathrm{RV}$ from
\begin{equation}
p_\mathrm{RV} = \frac{\Phi_\mathrm{cluster}(v_\mathrm{r})}{\Phi_\mathrm{cluster}(v_\mathrm{r}) + \Phi_\mathrm{field}(v_\mathrm{r})}
\end{equation}
for all stars with measured radial velocities from the data sets mentioned above. The membership probability distribution (Fig.~\ref{fig:RVprob}) is strongly bi-modal. We chose to include all stars with $p_\mathrm{RV} > 0.5$ as members because the radial velocity members are somewhat spread out at the high probability end, while most of the non-members have membership probabilities near zero. The spread for the radial velocity members is independent of the stellar mass and the radial velocity uncertainties; hence it comes from field star interlopers, the intrinsic cluster dispersion, and undetected binaries. We note that with the chosen threshold some field star non-members are likely included in this sample. We find 819 probable radial velocity member. All membership probabilities together with other data are given in Table~\ref{tab:rv}.

\begin{landscape}
	\begin{table}
		\caption{Radial velocities and membership probabilities of stars in the NGC~3532 field.}
		\label{tab:rv}
		\centering
		\begin{tabular}{lllllllllllllll}
			\hline
			\hline
			CLHW & RAJ2000 & DEJ2000 & $v_r$ & $\Delta v_r$ & $p_\mathrm{RV}$ & PM & M & Class & $V$ & $(B-V)$ & $(V-I_c)$ & $(V-K_s)$ & Name & Ref.\\
			& ($\degr$) & ($\degr$) & (km\,s$^{-1}$) & (km\,s$^{-1}$) &  &  &  &  & (mag) & (mag) & (mag) & (mag) & & \\
			\hline
			316006 & 165.38629 & -58.66414 & 7.30 & 4.63 & 0.88 & y & m & \dots & 17.43 & 1.47 & 1.93 & 4.06 & \dots & 1\\
			316085 & 165.39042 & -58.45367 & 7.43 & 2.50 & 0.87 & y & m & \dots & 16.79 & 1.42 & 1.62 & 3.51 & \dots & 1\\
			314978 & 165.39146 & -59.01008 & 5.11 & 3.67 & 0.94 & y & m & \dots & 18.39 & 1.46 & 2.29 & 4.30 & \dots & 1\\
			\dots & & & & & & & & & & & & & & \\
			244323 & 166.03392 & -58.79339 & 4.08 & 1.92 & 0.92 & y & m & SM & 16.06 & 1.27 & 1.41 & 3.05 & ID 203438, CTIO 4\_12400 & 1, 2\\
			244398 & 166.03617 & -58.51553 & 6.91 & 2.43 & 0.91 & y & m & U & 16.16 & 1.26 & 1.43 & 3.04 & ID 102467, CTIO 3\_9428 & 1, 2\\
			244108 & 166.03654 & -58.68747 & 6.06 & 1.85 & 0.94 & y & m & SM & 16.28 & 1.33 & 1.50 & 3.24 & ID 102481, CTIO 3\_9384 & 1, 2\\
			\dots & & & & & & & & & & & & & & \\
			244333 & 166.03254 & -58.91892 & 27.68 & 1.97 & 0.00 & n & n & BLN & 15.87 & 1.12 & 1.21 & \dots & ID 203421, CTIO 4\_12409 & 2\\
			244134 & 166.03400 & -58.91925 & 12.08 & 2.54 & 0.00 & n & n & BLN & 15.47 & 1.22 & 1.30 & \dots & ID 203440, CTIO 4\_12376 & 2\\
			241907 & 166.05104 & -58.85297 & 4.30 & 0.43 & 0.93 & y & m & U & 16.40 & 1.45 & 1.75 & 3.65 & ID 203721, CTIO 4\_11972 & 2\\
			\dots & & & & & & & & & & & & & & \\
			\dots & 164.96744 & -59.57667 & 10.60 & \dots & 0.49 & \dots & n & \dots & 8.74 & -0.03 & \dots & \dots & HD 95413, G 2 & 3\\
			\dots & 164.97040 & -58.16675 & 9.40 & \dots & 0.67 & \dots & m & \dots & 8.91 & 0.05 & \dots & 0.25 & HD 95412, G 1 & 3\\
			\dots & 165.03093 & -58.34816 & 2.50 & \dots & 0.82 & \dots & m & \dots & 9.37 & 0.04 & \dots & 0.12 & HD 95448, G 3 & 3\\
			\dots & & & & & & & & & & & & & & \\
			315560 & 165.40133 & -58.40169 & 8.65 & 11.86 & 0.59 & \dots & m & \dots & 13.03 & 0.65 & 0.60 & 1.38 & \dots & 4\\
			312333 & 165.42154 & -58.99550 & -4.95 & 1.36 & 0.00 & n & n & \dots & 12.60 & 0.65 & 0.72 & 1.58 & \dots & 4\\
			313635 & 165.42483 & -58.38972 & 17.36 & 2.68 & 0.00 & n & n & \dots & 12.47 & 0.53 & 0.59 & 1.26 & \dots & 4\\
			\dots & & & & & & & & & & & & & & \\
			134898 & 166.71192 & -58.68503 & 4.17 & 0.37 & 0.94 & y & m & \dots & 15.63 & 1.15 & 1.27 & \dots & GES 11065085-5841061 & 5\\
			123294 & 166.77296 & -58.65750 & 4.59 & 0.37 & 0.95 & y & m & \dots & 16.34 & 1.34 & 1.51 & 3.17 & GES 11070549-5839270 & 5\\
			121200 & 166.78446 & -58.74417 & 5.56 & 0.37 & 0.93 & y & m & \dots & 15.78 & 1.19 & 1.32 & 2.84 & GES 11070825-5844390 & 5\\
			\dots & & & & & & & & & & & & & & \\
			223513 & 166.18767 & -58.70257 & 4.60 & \dots & 0.98 & \dots & m & BU & 8.29 & -0.08 & 0.10 & 0.03 & HD 96213, BDA 155 & 6\\
			216879 & 166.22983 & -58.74963 & 3.30 & \dots & 0.99 & y & m & SM & 8.17 & 0.00 & 0.02 & 0.05 & HD 96227, BDA 40 & 6\\
			215077 & 166.24179 & -58.79508 & 2.00 & \dots & 0.97 & n & n & SN & 8.35 & -0.08 & 0.15 & 0.23 & HD 96246, BDA 50 & 6\\
			\dots & & & & & & & & & & & & & & \\
			241779 & 166.05317 & -58.72860 & -23.05 & 0.46 & 0.00 & n & n & SN & 7.91 & 0.56 & 0.79 & 1.71 & HD 96122, BDA 273 & 7\\
			301535 & 165.53463 & -59.04061 & 3.64 & 10.06 & 0.89 & y & m & \dots & 11.97 & 0.46 & 0.59 & 1.25 & RAVE J110208.3-590226 & 8\\
			275139 & 165.79721 & -58.18592 & 16.07 & 5.10 & 0.00 & y & n & \dots & 10.87 & 0.29 & 0.28 & 0.67 & RAVE J110311.4-581109 & 8\\
			271266 & 165.82500 & -58.48844 & 4.45 & 2.71 & 0.93 & y & m & \dots & 11.33 & 0.32 & 0.38 & 0.82 & RAVE J110318.0-582918 & 8\\
			
		\end{tabular}
		\tablebib{(1) This work, AAO observations; (2) This work, CTIO observations; (3) \cite{1981A&A....99..155G}; (4) \emph{Gaia} DR2 \citep{2018A&A...616A...1G}; (5) \emph{Gaia} ESO Survey DR2; (6) \cite{2002AJ....123.3318G}; (7) \cite{2008A&A...485..303M}; (8) RAVE survey DR5 \citep{2017AJ....153...75K}}
		\tablefoot{The full table is available at the CDS. \textit{CLHW}: ID from C11; \textit{PM}: proper motions member (y/n); \textit{M}: cluster membership from (mean) radial velocity, spectroscopic analysis, and proper motions (m) member, (n) non-member; \textit{Class}: Membership class for stars with multiple measurements: B: binary, G: background giant, L: likely, M: member, N: non-member, P: photometric (only non-members), S: single, U: unknown multiplicity; $V$ and $(B-V)$ photoelectric magnitudes \citep{1980A&AS...39...11F, 1982AJ.....87.1390W} if available, otherwise from C11, or from Tycho-2 \citep{2000A&A...355L..27H}, $I_c$: C11; $K_s$: 2MASS; \textit{Name}: ID: this work, CTIO: this work CTIO observations, G: \cite{1981A&A....99..155G}, GES: \emph{Gaia} ESO Survey, BDA: \cite{1992BICDS..40..115M}. The \emph{Gaia} DR2 IDs can be found in Table~\ref{tab:photomem}.}
		
	\end{table}
\end{landscape}

\subsection{Comparison with prior work}
\label{sec:rvprior}

Few prior radial velocity measurements are available in the literature. The first study of radial velocities in NGC~3532, based on objective prism spectra, was presented in two papers by \cite{1980A&AS...41..245G, 1981A&A....99..155G}. The author measured relative radial velocities for 84 stars and found a mean cluster radial velocity of 4.6\,km\,$\mathrm{s}^{-1}$ (relative to a zero point based on the radial velocity of other stars). In the study of \cite{2002AJ....123.3318G} some of the same stars were observed and the authors estimated the zero point for the Gieseking sample to be 2.4\,km\,$\mathrm{s}^{-1}$, and thus a mean cluster radial velocity of 2.2\,km\,$\mathrm{s}^{-1}$ relative to the barycentre of the solar system.

Independent of the zero point estimate, \cite{2002AJ....123.3318G} give a mean cluster velocity of $3.4\pm0.3$\,km\,$\mathrm{s}^{-1}$ from observations of 21 mainly giant stars. The radial velocity study by \cite{2008A&A...485..303M} targeted eight red giants in NGC~3532 and obtained a mean cluster velocity of $4.3\pm{0.34}$\,km\,$\mathrm{s}^{-1}$.

First results from the GES were presented in \cite{2016A&A...591A..37J}. This study included only two stars in NGC~3532 although many more were observed and are provided in the data releases. For the two stars in \cite{2016A&A...591A..37J} radial velocities of 3.8\,km\,$\mathrm{s}^{-1}$ and 5.8\,km\,$\mathrm{s}^{-1}$ were found. The full sample of GES stars was analysed in the joint data set above.

Our mean radial velocity of $v_\mathrm{r}=5.43\pm{0.04}$\,km\,s$^{-1}$ is somewhat different from the cluster velocities determined by \cite{2002AJ....123.3318G} and \cite{2008A&A...485..303M}. Those two studies targeted the massive stars in the cluster, while our sample includes exclusively cooler dwarf stars. Because of the mean velocity differences and the disjunct data sets we have not fitted the whole cluster with a single model. Instead, we refrained from recalculating the membership probability and use the values given by \cite{2002AJ....123.3318G} and \cite{2008A&A...485..303M}. Additionally, we reanalysed the data of \cite{1981A&A....99..155G} with a two-component Gaussian mixture model. We find the parameters compatible with the widths and positions of the Gaussians estimated by \cite{1981A&A....99..155G} and derive membership probabilities. Later, we define the cluster sequence from the full set of probable members.

\subsection{Repeated radial velocity observations}
\label{sec:multiobs}

The aim of our observing campaign at CTIO was to discern between the single and binary members of NGC~3532. Within three years we obtained up to nine radial velocity data points per star and are able to find candidate single members. Later at AAO, 145 of the CTIO targets were re-observed, extending the time-series to a baseline time-scale of a decade. Single members are easily confirmed and even long-term radial velocity variations may be found in these data.

We have obtained more than one observation for 334 of the 372 observed stars and at least three for 276 stars. We analysed the radial velocity time-series by visual inspection and with the $e/i$-statistic \citep{2008AJ....135.2264G}, where $e$ is the standard deviation of the radial velocities and $i$ the precision. The $e/i$ threshold used is 4. Among the stars with multiple observations we find 41 radial velocity variables, of which 27 can be identified as likely binaries with oscillating radial velocities. The remaining 14 have too few data points to draw firm conclusions. We have not attempted to determine a $\gamma$ velocity for the likely binaries because the time-series contain too few data points at present to solve the orbits.

We have assigned labels to the data in Table~\ref{tab:rv} on the basis of these time-series. They include \textit{B} for binary, \textit{S} for single star and \textit{U} for unknown variability, the latter the result of having to too few data points\footnote{Labels were only assigned to stars with multiple observations from our own data sets and known binaries from other data sets.}. For all stars with multiple measurements we calculated a mean radial velocity which we use for our analysis.

Having obtained only a single observation at AAO we do not know whether stars in this data set have constant radial velocities. Despite this uncertainty we have included all stars with a single measurement from the AAO into the determination of the membership probabilities. Consequently, some of these stars could potentially be binaries, and their observed orbital velocities could disguise their true membership status. Additionally small radial velocity variability could potentially widen the radial velocity distribution.

Assuming the Gaussian two-component model to be correct, we can also deduce membership for the radial velocity variables. For each variable star we calculate the mean radial velocity and apply the trained model to it. We publish the membership probabilities determined through this method as is, but note that they should be used with care\footnote{The stars in question have a class label of either \textit{B} or \textit{U}.}. 

\section{Additional stellar parameters from the spectra}
\label{sec:stellpar}

\begin{figure*}
	\includegraphics[width=\textwidth]{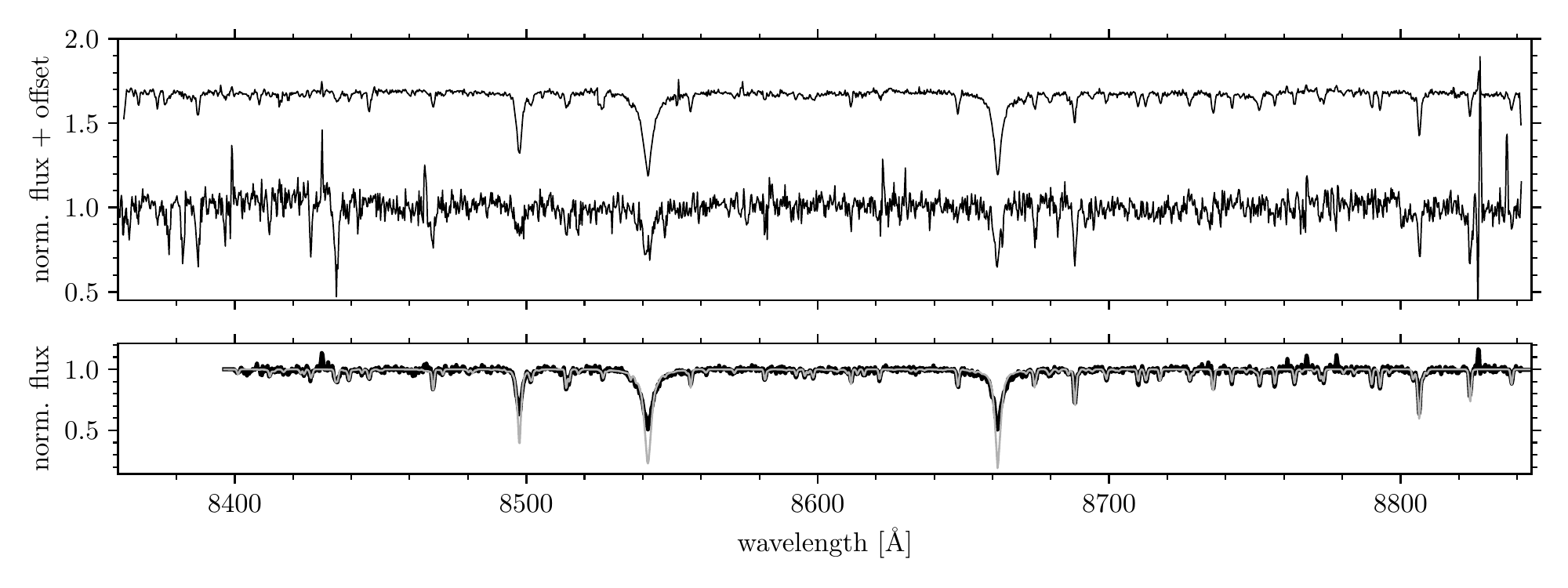}
	\caption{Spectra for three stars of our sample observed with AAOmega. \textit{Top:} Examples for a spectrum with a high SNR (SNR$\sim150$, $I_c=12$, upper) and a spectrum with a low SNR (SNR$\sim15$, $I_c=17$, lower). \textit{Bottom:} spectrum of a solar-like radial velocity member (black) with the best fit by \textsc{SP\_Ace} shown in light grey.}
	\label{fig:spectra}
\end{figure*}

To obtain information beyond the radial velocities from the AAOmega spectra (see Fig.~\ref{fig:spectra} for three examples) we used the software \textsc{SP\_Ace} (Stellar Parameter And Chemical abundances Estimator, \citealt{2016A&A...587A...2B}, version 1.1) which estimates the stellar parameters based on polynomial fits to the equivalent widths of several spectral lines. The parameters for those fits are stored in the General Curve-Of-Growth library (\textsc{GCOG}) which is included in \textsc{SP\_Ace}. \textsc{GCOG} is based on calibrated\footnote{\cite{2016A&A...587A...2B} use the Sun, Arcturus, Procyon, $\epsilon$ Eri, and $\epsilon$ Vir for calibration.} oscillator strengths from the line list of the Vienna Atomic Line Database (VALD, \citealt{1999A&AS..138..119K}) and 1D atmospheric models synthesized from ATLAS9 \citep{1997A&A...318..841C} under LTE assumptions.

The best-fitting grid point from the \textsc{GCOG} library is determined through an iterative Levenberg-Marquardt minimization routine. To find the stellar parameters and abundances the equivalent width of that grid point is used. \textsc{SP\_Ace} can also output a synthetic spectrum based on those parameters. One illustrative example is shown in Fig.~\ref{fig:spectra} (bottom panel). For this spectrum \textsc{SP\_Ace} achieved $\chi^2=1.06$. This fit is representative of our analysis where more than 90 per cent of the successful fits have $\chi^2 < 1.2$. 

The grid of \textsc{SP\_Ace} is restricted to $T_\mathrm{eff} > 3600\,\mathrm{K}$, a fact which implies that parameters cannot be estimated for the coolest stars in our sample. In any case, stars in our sample with $T_\mathrm{eff} < 4000\,\mathrm{K}$ do not have reliable measurements due to their faintness and the low SNRs of their spectra. 

The estimated stellar parameters include effective temperature ($T_\mathrm{eff}$), gravity ($\log g$), and various abundances, particularly the iron abundance ([Fe/H]) relative to the Sun. [Fe/H] can be measured from $\sim$~60 lines in the spectral range, where the exact number depends on the SNR. The typical uncertainties arising in the analysis of the AAOmega spectra are $\Delta T_\mathrm{eff}=100$\,K, $\Delta\log g=0.23$, and $\Delta\mathrm{[Fe/H]}=0.06$. Apart from the figures presented in this work (e.g. Fig.~\ref{fig:HRD}) we have tested the results of \textsc{SP\_Ace} for consistency with additional checks. For example we have checked that $T_\mathrm{eff}$ plotted against colour shows a clear cluster sequence.

We were able to estimate atmospheric parameters for a total of 355 stars in our sample. The individual values for each star are given in Table~\ref{tab:stellpar}.

\begin{table*}
	\caption{Stellar parameters obtained from the spectra using \textsc{SP\_Ace}.}
	\centering
	\label{tab:stellpar}
	\begin{tabular}{lllllllllll}
	\hline
	\hline
	CLHW & ID & RAJ2000 & DEJ2000 & $p_\mathrm{RV}$ & $T_\mathrm{eff}$ & $\Delta T_\mathrm{eff}$ & $\log g$ & $\Delta \log g$ &  [Fe/H] & $\Delta \mathrm{[Fe/H]}$ \\
	& & ($\degr$) & ($\degr$) & & (K) & (K) &  &  &  & \\
	\hline
		310132 & 800249 & 165.44771 & -58.87789 & 0.94 & 5471 & 133 & 4.41 & 0.23 & -0.15 & 0.07\\
		309783 & 800325 & 165.45367 & -58.80531 & 0.74 & 5264 & 107 & 4.72 & 0.19 & -0.26 & 0.06\\
		300827 & 802057 & 165.54779 & -58.81253 & 0.89 & 4684 & 127 & 4.50 & 0.24 & 0.02 & 0.07\\
		299782 & 802295 & 165.55558 & -58.90439 & 0.94 & 5464 & 164 & 4.24 & 0.36 & -0.03 & 0.08\\
		294532 & 803729 & 165.61046 & -58.64156 & 0.00 & 6495 & 479 & 3.72 & 0.67 & 0.10 & 0.11\\
		291918 & 804397 & 165.62996 & -58.85247 & 0.00 & 6153 & 178 & 4.04 & 0.33 & 0.02 & 0.04\\
	\dots
\end{tabular}
\tablefoot{The full table is available at the CDS. \textit{CLHW}: ID from C11; \textit{ID}: ID this work; \textit{RAJ2000}, \textit{DEJ2000}: from C11.}
\end{table*}

\subsection{Surface gravity and effective temperature}
\begin{figure}
	\includegraphics[width=\columnwidth]{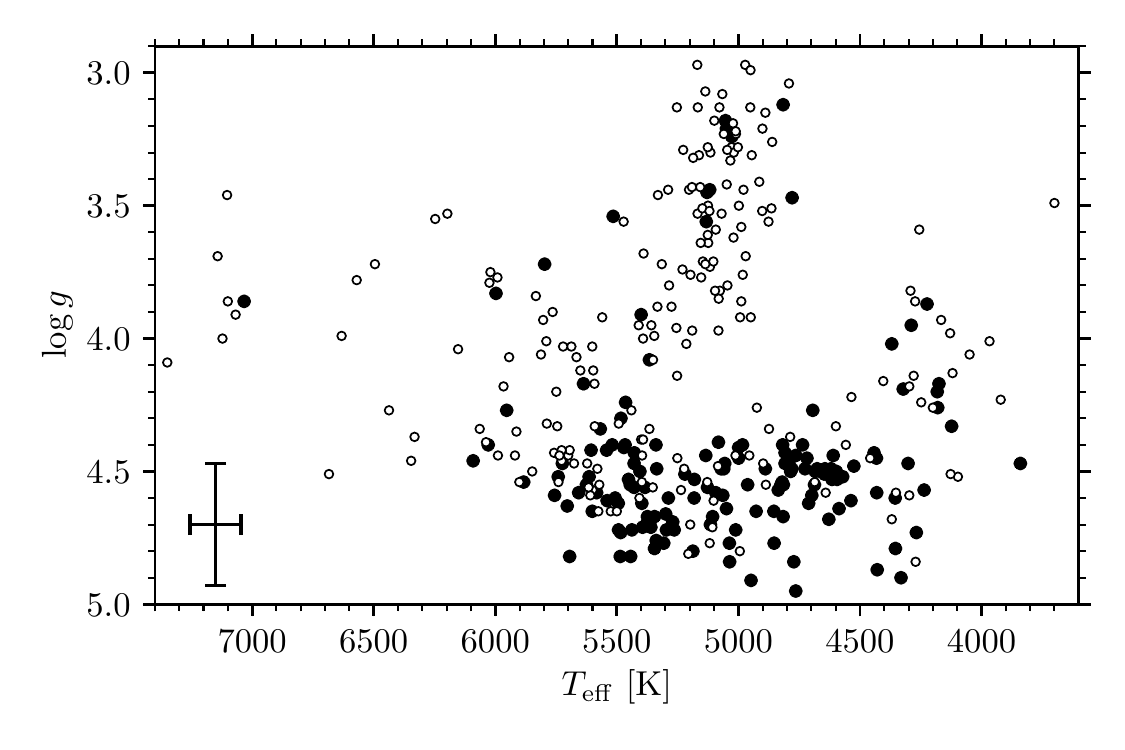}
	\caption{Gravity ($\log g$) against effective temperature ($T_\mathrm{eff}$) for stars observed spectroscopically with AAOmega. Filled circles are radial velocity members while the unfilled ones are non-members. Some radial velocity members are found among the background giant population. In the lower left we show the typical errors.}
	\label{fig:HRD}
\end{figure}

The derived $\log g$ and $T_\mathrm{eff}$ values are displayed in Fig~\ref{fig:HRD}, in combination with our derived membership status. We find 93 (out of 233) dwarf stars ($\log g > 4.0$) with obtained surface gravity to be radial velocity non-members. This shows that in the field of NGC~3532 there is a significant field contamination from foreground stars.

The photometric membership list is also contaminated by a population of (background) giant stars which we were unable to remove through the colour-colour diagrams. In the CMDs (Fig.~\ref{fig:cmd}) a background component crosses the cluster sequence at $(B-V)=1.3$ which matches the reddened stellar parameters of the giant stars ($T_\mathrm{eff}\approx5300$\,K and $(B-V)_0\approx 0.9$). Among the giants we find several radial velocity members which we can now remove from the sample of cluster members in order to obtain a cleaner cluster sequence, reducing the number of probable members to 804. We therefore see that obtaining the additional parameters $\log g$ and $T_\mathrm{eff}$ from the spectra can help to clean up the sample of cluster members beyond the radial velocity information.

With the additional information about background giant stars we recalculated the radial velocity distribution but found only minor differences from the initial distribution. Hence, we do not update the membership probability and continue with the previously calculated values while marking the background giants as non-members.

\subsection{Abundances of NGC~3532}
\label{sec:meh}

A number of prior estimates for the metallicity of NGC~3532 are available in the literature. The first photometrically determined metallicity was presented by \cite{1988Obs...108..218C} $\mathrm{[Fe/H]}=0.08\pm0.08$ while the first spectroscopic study by \cite{1994ApJS...91..309L} found $\mathrm{[Fe/H]}= 0.07\pm0.06$ for the cluster giants.

A photometric estimate from DDO photometry by \cite{1995AJ....110.2813P} gave $\mathrm{[Fe/H]}=-0.10\pm0.09$, also for red giants. The same data and additional $UBV$ photometry were used by \cite{1997AJ....114.2556T} who reported $\mathrm{[Fe/H]}=-0.02\pm0.09$ for the same set of stars.

\cite{2000ASPC..198..225G} recalibrated all the prior studies and found $\mathrm{[Fe/H]}=0.02\pm0.06$. \cite{2001A&A...373..159C} listed eleven giant stars, of which six are radial velocity members of NGC~3532 with a mean metallicity of $\mathrm{[M/H]}=0.06$. \cite{2012A&A...538A.151S} compared different line lists and found a range from $\mathrm{[Fe/H]}=-0.06\pm0.07$ to $\mathrm{[Fe/H]}=0.03\pm0.03$ for giants in NGC~3532.

The RAVE survey was the first to publish metallicities for dwarf stars in NGC~3532. From seven stars \cite{2014A&A...562A..54C} found $\mathrm{[M/H]}=-0.021\pm0.057$. From high-quality spectra of four stars \cite{2016A&A...585A.150N} found $\mathrm{[Fe/H]}=0.00\pm0.07$ and a photometric  metallicity estimate by \cite{netopil} gave $\mathrm{[Fe/H]}=0.0\pm0.06$ for giants stars and $\mathrm{[Fe/H]}=0.02\pm0.12$ from 49 dwarfs. The GES data release 2 unfortunately does not have metallicity estimates for NGC~3532.

All of these suggest a cluster with near-solar metallicity value, with an error in the range of 0.1\,dex. We note that most prior estimates were made for a small sample of giant members and not the dwarf stars in NGC~3532. With our observations of solar-like stars from NGC~3532 we now have a large sample of dwarf stars at hand.

\textsc{SP\_Ace} is capable of measuring several different abundances from the spectra. Because of the relatively low SNR we chose to use only the [Fe/H] values in this study. The mean cluster iron abundance, measured from the 139 radial velocity dwarf members, is $\mathrm{[Fe/H]}=-0.07\pm0.10$.

\begin{figure}
	\includegraphics[width=\columnwidth]{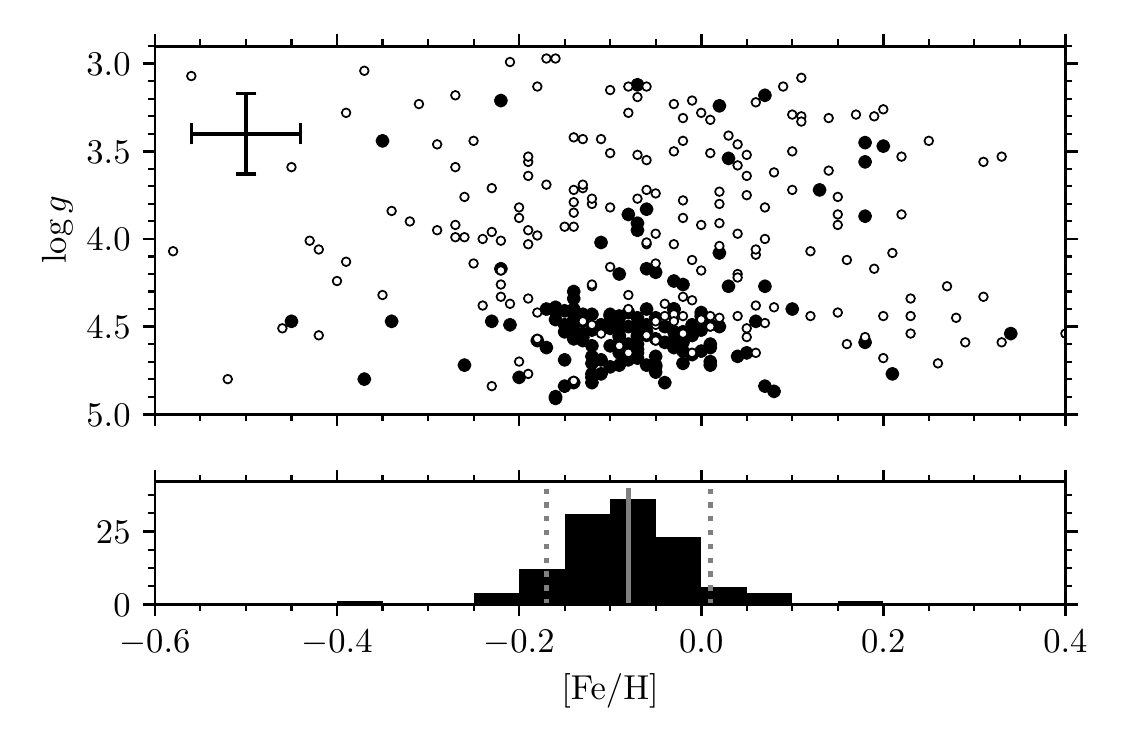}
	\caption{\textit{Top:} Gravity ($\log g$) dependence on the iron abundance ([Fe/H]) relative to the Sun for radial velocity members (filled) and non-members (unfilled). The typical individual measurement error is given in the upper left corner. \textit{Bottom:} Histogram of the iron abundance, including only the cluster dwarfs ($\log g > 4$). The median value of the distribution (solid line) is $\mathrm{[Fe/H]}_\mathrm{cluster}=-0.07\pm0.1$. The dashed lines show the standard deviation.}
	\label{fig:MeH}
\end{figure}

The metallicity distributions for both the members and non-members are shown in Fig.~\ref{fig:MeH}, including a histogram for the cluster dwarf members. Although a solar-like metallicity for the open cluster is within the uncertainties, the determined mean metallicity is slightly sub-solar.

\section{Astrometric data from \emph{Gaia}}
With the information from \emph{Gaia} DR2 \citep{2018A&A...616A...1G} becoming available after our observation, we have the possibility of verifying the proposed members with precise astrometric parameters. We first use the proper motions, and analyse whether they are suitable for refining the cluster sequence. Second, we turn to the parallaxes to calculate the cluster distance.

\subsection{Proper motions of cluster members}
\label{sec:pm}

\begin{figure*}
	\includegraphics[width=\columnwidth]{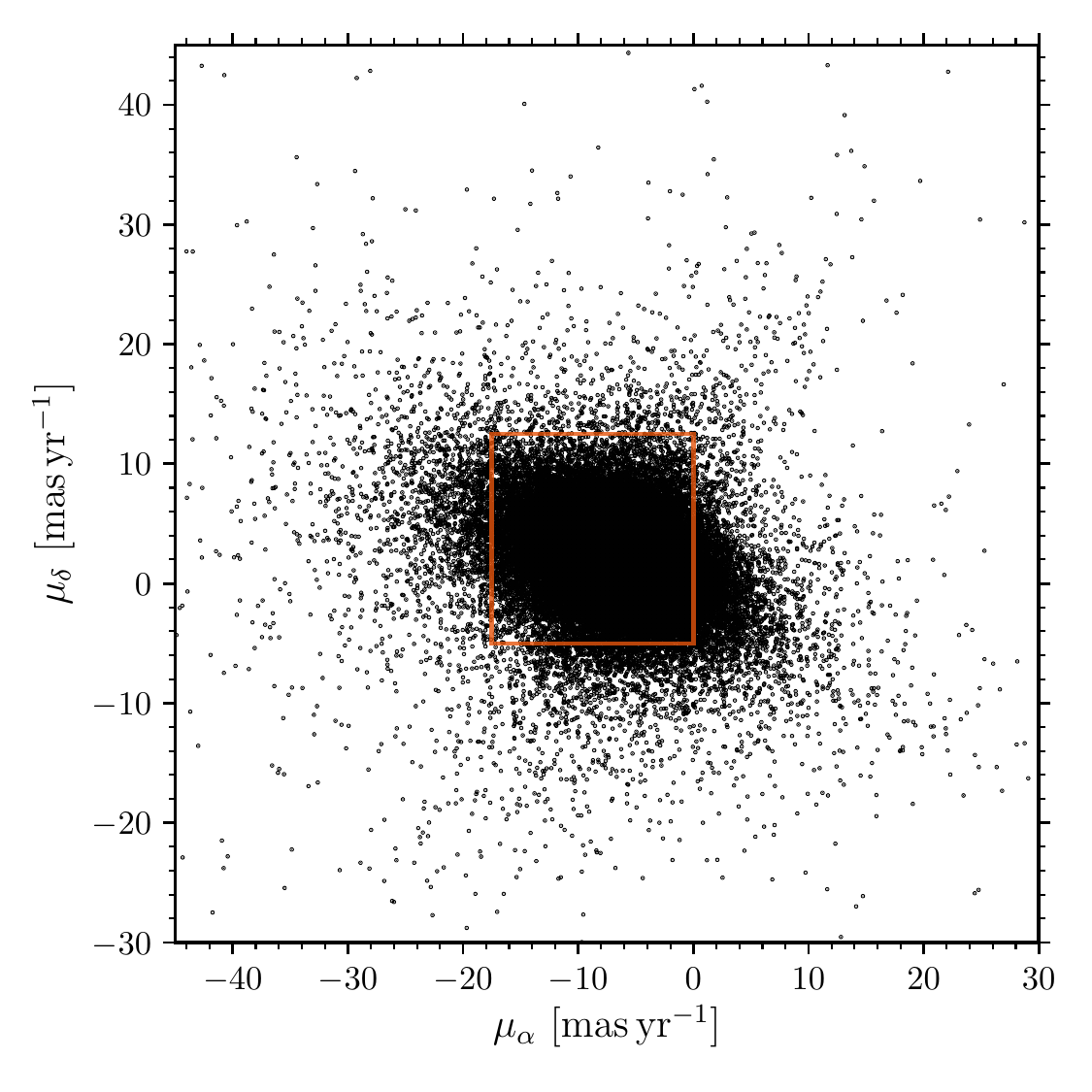}
	\includegraphics[width=\columnwidth]{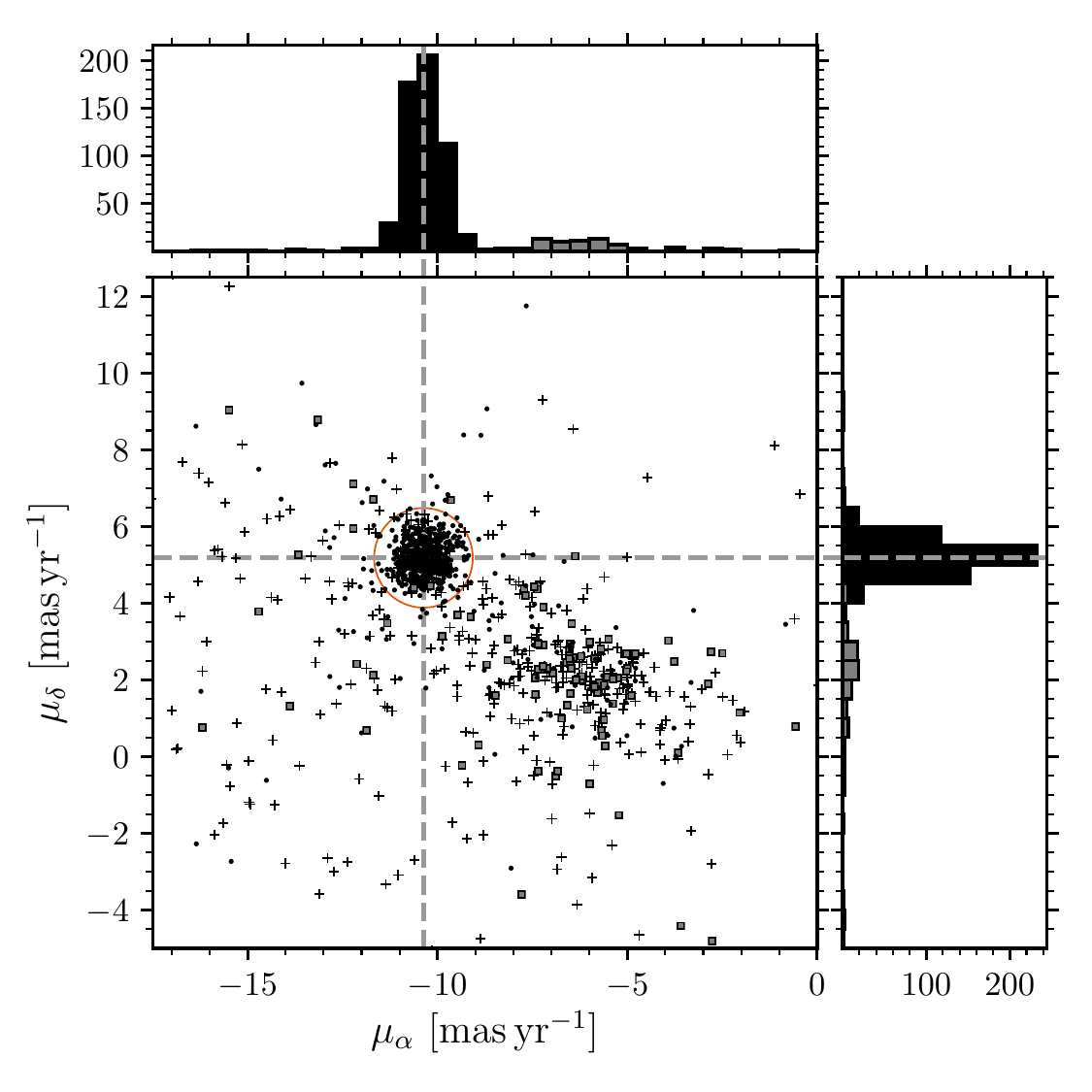}
	\caption{Proper motion diagram of the NGC~3532 field. \textit{Left:} All stars common to the catalogues of C11 and Gaia DR2. In this plot no difference between the proper motions of the field and cluster stars is visible. \textit{Right:} Same as left but zooming in on the region outlined on the left, with the sample limited to the spectroscopically observed stars. Here the cluster (black dots) is distinguishable from the field (crosses and grey squares). We have separated the field into the background giant component (grey squares) and the foreground component(crosses). The histograms show the distribution of radial velocity members (black) and background giant stars (grey). The dashed lines give the centre of the distribution of the members ($\mu_\alpha = -10.37\pm0.16$\,mas\,yr$^{-1}$, $\mu_\delta=5.18\pm0.08$\,mas\,yr$^{-1}$) and the  circle defines the region of proper motion members.}
	\label{fig:pm}
\end{figure*}

Knowing the radial velocity members of NGC~3532, we can extract additional information from the proper motions. From the precise proper motions provided in \emph{Gaia} DR2, with uncertainties $\lesssim0.1$\,mas\,yr$^{-1}$, we can derive clean cluster membership. In Fig.\ref{fig:pm} we plot in the left panel all stars in the field. The open cluster and the field are inseparable, due to the large number of stars in the area. Even in a density plot the separation is not easy to achieve. We show in the right panel only the proper motions of the photometric cluster members for which we have obtained radial velocities. From the analysis of the spectra we can split these stars into three groups: radial velocity members, non-members and background giant non-members. 

The most obvious feature is the tight clustering of radial velocity members around $\mu_\alpha = -10.37\pm0.16$\,mas\,yr$^{-1}$, $\mu_\delta=5.18\pm0.08$\,mas\,yr$^{-1}$ marked in Fig.~\ref{fig:pm} with dashed lines. This centre is the mean proper motion of the radial velocity members in this clustering with the uncertainty given by the standard deviation. The proper motion found for NGC~3532 is marginally in agreement for $\mu_\alpha$ with the published proper motion from both Hipparcos ($\mu_\alpha = -10.84\pm0.38$\,mas\,yr$^{-1}$, $\mu_\delta=5.26\pm0.37$ \citealt{1999A&A...345..471R}) and \emph{Gaia}/TGAS ($\mu_\alpha = -10.54\pm0.03$\,mas\,yr$^{-1}$, $\mu_\delta=5.19\pm0.04$\,mas\,yr$^{-1}$, \citealt{2017A&A...601A..19G}). Our value though is (unsurprisingly) confirmed by \cite{2018A&A...616A..10G}, who found a very similar proper motion for NGC~3532 ($\mu_\alpha = -10.3790\pm0.0079$\,mas\,yr$^{-1}$, $\mu_\delta=5.1958\pm0.0079$\,mas\,yr$^{-1}$) also from \emph{Gaia} DR2.

Although the centre of the cluster proper motions is well-defined, some radial velocity members are scattered over the whole proper motion plane. We chose to include all stars within a radius of 1.3\,mas\,yr$^{-1}$ around the cluster centre in proper motion space as proper motion members (circle in Fig.~\ref{fig:pm}) because out to this radius the number of proper motion members increases non-linearly with radius. For larger radii the linear increase of proper motion members observed can be attributed to field stars, while the non-linear increase is due to the clustering of stars. The radial velocity members follow the same pattern, a fact which gives us confidence about the chosen cut-off value. Among the included stars we find 58 radial velocity non-members, each of which could potentially a be binary (because we have only obtained a single radial velocity measurement for these stars). However, at least six of the proper motion members have multiple radial velocity observations and are classified as single non-members, a fact which demonstrates that background contamination remains in the proper motion membership. After all proper motion membership, like that from radial velocities, is probabilistic. The proper motion membership is given in a column in Tables~\ref{tab:photomem} and \ref{tab:rv}. By jointly demanding both radial velocity and proper motion membership, we define an exclusive set of bona fide cluster members which are listed as \textit{m} in the joint membership column (\textit{M}) in Table~\ref{tab:photomem}.

In Fig.\ref{fig:pm} we also highlight the background giant stars from our photometric membership. Those stars might be assumed to define the ``field''. Their proper motion is centred on $\mu_\alpha \approx -5$\,mas\,yr$^{-1}$, $\mu_\delta \approx 5$\,mas\,yr$^{-1}$. In the absence of better information C11 assumed $\mu_\alpha = 0$\,mas\,yr$^{-1}$, $\mu_\delta=0$\,mas\,yr$^{-1}$. The giant stars show a wider distribution in the proper motion plane than the compact cluster. Noticeable in this context are the stars identified as radial velocity members but with a low $\log g$. Among those giants with the same radial velocity as the open cluster we find some which even have very similar proper motions. This is a consequence of the low proper motion of NGC~3532 combined with the wide distribution of proper motions of field stars. This validates our cautious approach to assignment of cluster membership based on photometry, radial velocity, or proper motion alone. Sometimes spectral analysis is the only way to uncover false-positives.

\subsection{Astrometric distance to NGC~3532}
\label{sec:dist}

\emph{Gaia} DR2 has provided parallaxes for most of the stars in our sample of cluster stars. We have included all sources with positive parallaxes and relative uncertainties less than 10 per cent. In our case we use the provided parallaxes regardless of the known correlations on small scales because we use a simple mean to estimate the distance \citep{2018AJ....156...58B}. Other known systematics of the \emph{Gaia} DR2 include a small zero-point offset of the parallaxes which is dependent on the sky position, and dependencies of the parallax measurements on colour and magnitude \citep{2018A&A...616A...2L}. All effects are superseded by the uncertainties and the spread of parallaxes in the cluster. Additional, presently unknown, global and local systematics might be included in the data but should not be relevant to our aim of determining the distance to the open cluster.

In order to compute the mean parallax of the cluster we chose to use only the stars which are proper motion and radial velocity members. From this set of stars we find $\varpi=2.068\pm0.139$\,mas ($484^{+35}_{-30}$\,pc), a value which is beyond that from the Hipparcos results ($406^{+76}_{-56}$\,pc). Our distance agrees with the 484\,pc from the \cite{2018A&A...616A..10G}. We note that this distance is simply the mean distance of the members and not the centre of mass distance. This value certainly agrees with the isochrone fitting distance of C11 ($492^{+12}_{-11}$\,pc), reducing the tension between the prior astrometric distance and theoretical models.

\begin{figure}
	\includegraphics[width=\columnwidth]{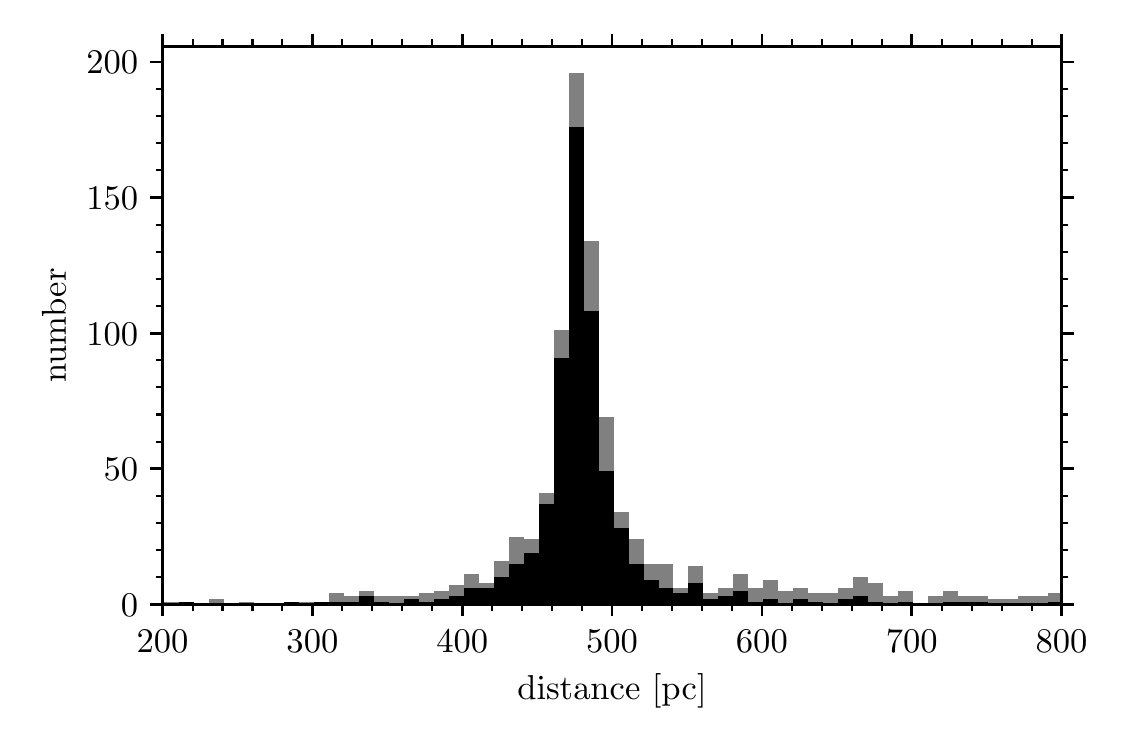}
	\caption{Distribution of Gaia distances for stars with measured radial velocities. The distribution of radial velocity members is shown in black, while the background (grey) includes all stars. The mean cluster distance (inferred from the parallaxes) is $484^{+35}_{-30}$\,pc. We only show a distance range near the open cluster.}
	\label{fig:distances}
\end{figure}

In Fig.~\ref{fig:distances} we show the distribution of distances to illustrate the size of NGC~3532. For this figure we use the individual distances provided by \cite{2018AJ....156...58B}. The peak indicating the cluster position is prominent, but the distribution is very wide for an open cluster. We find a number of radial velocity members which are 80\,pc closer or more distant to the Sun than the cluster centre in this distribution. Those stars are not only radial velocity--but also \emph{Gaia} DR2 proper motion members. The Pleiades for comparison have a radius of $\sim10$\,pc, using the members from \cite{2018A&A...616A..10G}. NGC~3532 is probably intrinsically more extended or has tidal tails like the Hyades \citep{2018arXiv181104931M, 2018arXiv181103845R}.  In view of this issue we do not use the distance as a membership criterion.

\section{Final Membership, cluster sequence, and isochrones}
\label{sec:clseq}

With the above-presented analysis of the radial velocities, stellar parameters, and proper motions we have found an exclusive and final set of 660 cluster members among 1613 spectroscopically observed photometric members\footnote{This number includes all observations of unique stars presented in this work and the data from the literature.}. By \textit{exclusive} we mean that all relevant criteria (photometry, radial velocity, and proper motion) are fulfilled without exceptions, creating a high quality data set for further studies of NGC~3532. These are the members shown in the following CMDs in this paper.

We present the exclusive members under column \textit{M} in Table~\ref{tab:rv}, together with all other radial velocity data, including both data from this study and the literature. We give the position of the star, its radial velocity, membership probability, the proper motion membership, the $V$ magnitude, and colours in $(B-V)$, $(V-I_c)$, and $(V-K_s)$. The $V$ magnitude and the $(B-V)$ colour are photoelectric measurements from \cite{1980A&AS...39...11F} and \cite{1982AJ.....87.1390W} where available, and CCD photometry from C11 otherwise. The column \textit{Ref.} gives the original publications for the radial velocity data. Furthermore, we included the identifiers from C11 for all stars that were observed in that study. Additionally, we give the identifiers assigned to the stars by the various studies including our own numbers and the labels for stars with multiple observations from Sec~\ref{sec:multiobs}.

In the rest of this paper we will work with this set of 660 exclusive members. First, we construct an empirical cluster sequence, tracing the locus of the open cluster in several colour-magnitude diagrams. Next, we compare various isochrone models to the observed cluster stars and finally, we estimate the total number of stars in this open cluster.

\subsection{Empirical cluster sequence}

A defining characteristic of an open cluster is the single-star main sequence along which almost all the members are distributed. With the set of exclusive members we can trace this main sequence in a CMD and construct a cluster sequence from the locus of the stars.

\begin{figure}
	\includegraphics[width=\columnwidth]{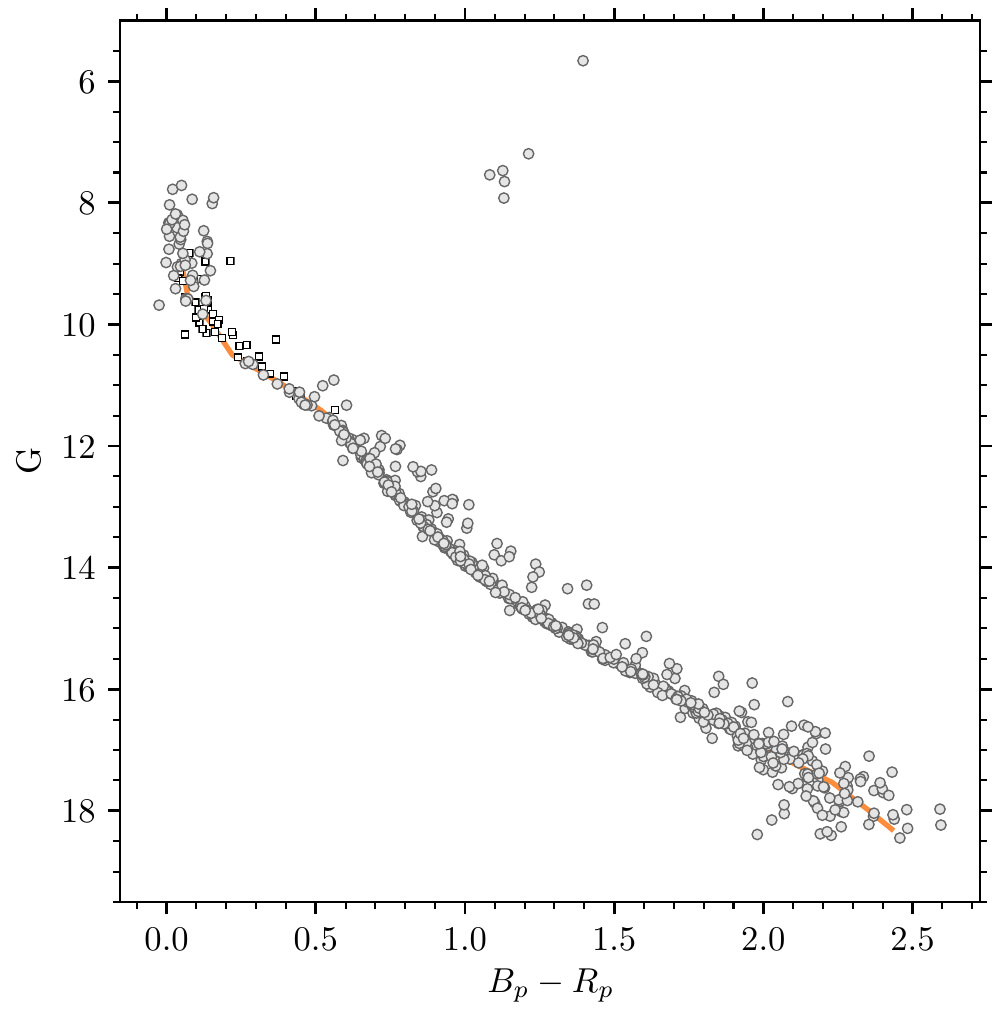}
	\caption{Colour-magnitude diagram of the exclusive set of members from \emph{Gaia} photometry. In the background we show our empirical cluster sequence (see. Table~\ref{tab:empiso}).}
	\label{fig:GaiaSeq}
\end{figure}

In order to accomplish this we plot CMDs in $(B-V)$, $(V-R_c)$, $(V-I_c)$, and $(V-K_s)$ against $V$. In addition we make use of the excellent photometry provided by \emph{Gaia} DR2. The [$(B_p-R_p)$, $G$] CMD is shown in Fig.~\ref{fig:GaiaSeq}. In each of these CMDs we trace the cluster sequence manually and tabulate the colours for fixed magnitudes. In order to improve the accuracy we use colour-colour diagrams and corrected small deviations. A magnitude-magnitude diagram helped us to find a smooth transformation between $V$ and $G$. In Fig.~\ref{fig:GaiaSeq} we present this final cluster sequence in the \emph{Gaia} photometric system. We display the data (exclusive cluster members) on top of the cluster sequence to emphasise the good match.

In Table~\ref{tab:empiso} we present the data for the cluster sequence in two photometric passbands and five different photometric colours. For the two photometric passbands we converted the magnitudes to absolute values using the \emph{Gaia} distance. All data were de-reddened. We use the coefficients from \cite{1968nim..book..167J} and for the \emph{Gaia} photometry we applied the relations from \cite{2010A&A...523A..48J}.

This cluster sequence can be used in future for comparison with other open clusters. Our sequence reaches down to $\sim0.35\,\mathrm{M}_\sun$, equivalent to a spectral type of M3V. We placed the brighter end of our sequence near the turn-off which, for NGC~3532 is populated by A0V stars  with $\mathrm{M}\sim2.6\,\mathrm{M}_\sun$.

\begin{table*}
	\caption{Empirical cluster sequence of NGC~3532 with $M_V$ and $M_G$ as well as several photometric colours. All data was de-reddened and shifted with the \emph{Gaia} distance.}
	\centering
	\label{tab:empiso}
	\begin{tabular}{lllllll}
		\hline
		\hline
		$M_V$ & $M_G$ & $(B-V)_0$ & $(V-R_c)_0$ & $(B_p-R_p)_0$ & $(V-I_c)_0$ & $(V-K_s)_0$ \\
		\hline
		0.48 &   0.48 & -0.014 &  0.008 &  0.001 & -0.005 &  0.01 \\
		0.98 &   0.98 &  0.046 &  0.009 &  0.021 &  0.025 &  0.11 \\
		1.48 &   1.46 &  0.106 &  0.038 &  0.101 &  0.065 &  0.22 \\
		1.98 &   1.98 &  0.171 &  0.083 &  0.171 &  0.165 &  0.37 \\
		2.48 &   2.44 &  0.246 &  0.143 &  0.331 &  0.260 &  0.55 \\
		2.98 &   2.90 &  0.336 &  0.198 &  0.471 &  0.365 &  0.77 \\
		3.48 &   3.38 &  0.416 &  0.243 &  0.581 &  0.470 &  1.01 \\
		3.98 &   3.91 &  0.486 &  0.278 &  0.651 &  0.545 &  1.16 \\
		4.48 &   4.33 &  0.566 &  0.323 &  0.731 &  0.630 &  1.34 \\
		4.98 &   4.84 &  0.646 &  0.363 &  0.821 &  0.685 &  1.51 \\
		5.48 &   5.28 &  0.751 &  0.418 &  0.931 &  0.770 &  1.72 \\
		5.98 &   5.78 &  0.851 &  0.473 &  1.051 &  0.865 &  2.06 \\
		6.48 &   6.23 &  0.966 &  0.533 &  1.161 &  0.995 &  2.30 \\
		6.98 &   6.68 &  1.086 &  0.638 &  1.311 &  1.155 &  2.55 \\
		7.48 &   7.03 &  1.206 &  0.718 &  1.451 &  1.335 &  2.87 \\
		7.98 &   7.48 &  1.336 &  0.783 &  1.611 &  1.495 &  3.20 \\
		8.48 &   7.85 &  1.406 &  0.863 &  1.761 &  1.665 &  3.51 \\
		8.98 &   8.23 &  1.446 &  0.918 &  1.881 &  1.845 &  3.73 \\
		9.48 &   8.65 &  1.466 &  0.973 &  2.031 &  2.045 &  4.01 \\
		9.98 &   9.01 &  1.476 &  1.023 &  2.181 &  2.225 &  4.31 \\
		10.48 &  9.46 &  1.486 &  1.073 &  2.301 &  2.345 &  4.53 \\
		10.98 &  9.78 &  1.496 &  1.113 &  2.381 &  2.465 &  4.81 \\
		\hline
	\end{tabular}
\end{table*}

\subsection{Comparison with isochrones}
\label{sec:isochrone}

\begin{figure*}
	\includegraphics[width=\textwidth]{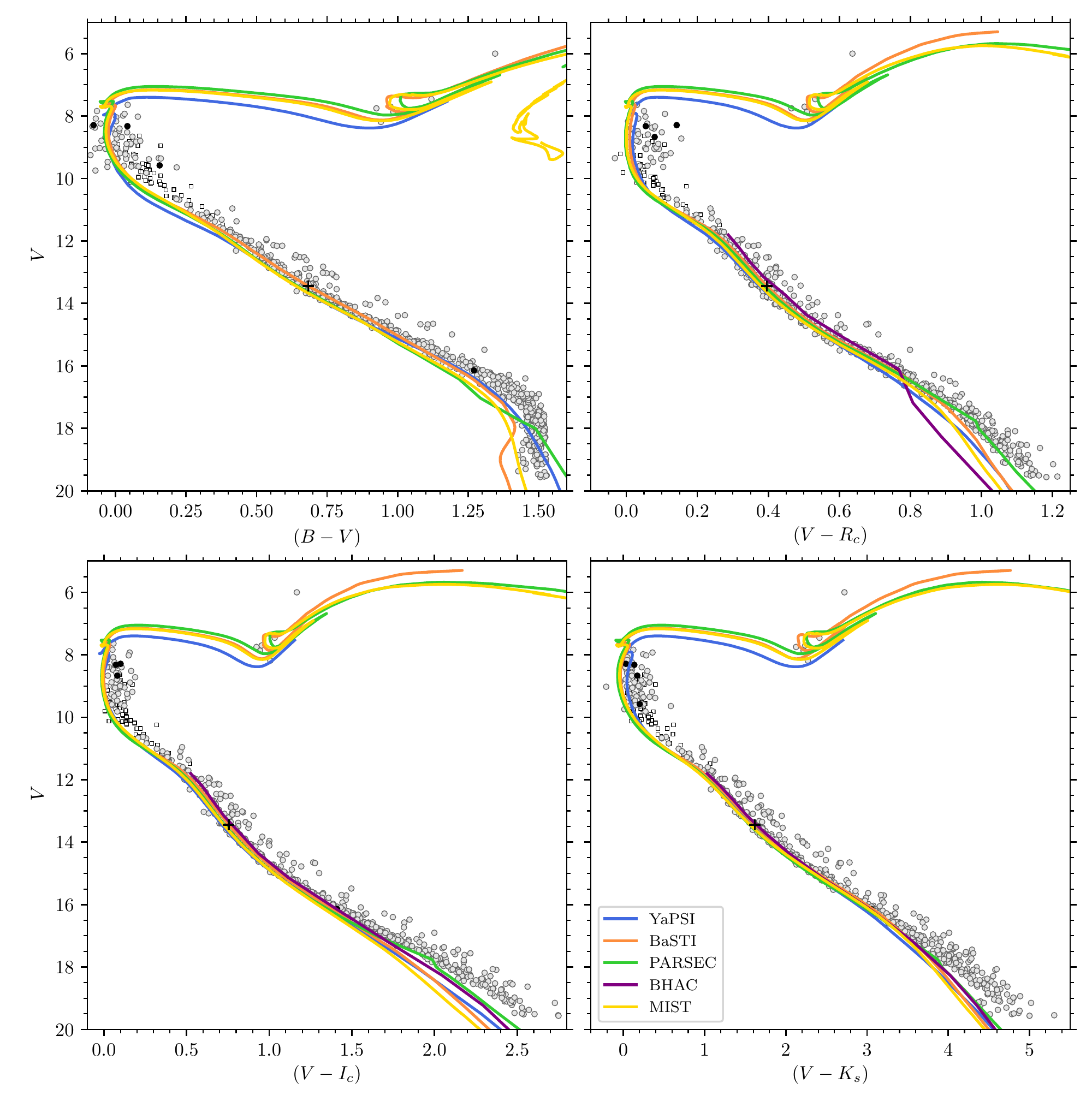}
	\caption{Radial velocity single cluster members (grey circles), (likely) binary members (black circles) and additional Gaia/TGAS proper motion members (white squares) of NGC~3532 in four colour-magnitude diagrams. For comparison with each other and the data we show the YaPSI (blue), PARSEC (green), BaSTI (orange), MIST (yellow), and BHAC (purple, not available in $(B-V)$) models. The solar-mass model for 300\,Myr is marked with a black plus sign.}
	\label{fig:isos}
\end{figure*}

\begin{figure*}
	\includegraphics[width=\textwidth]{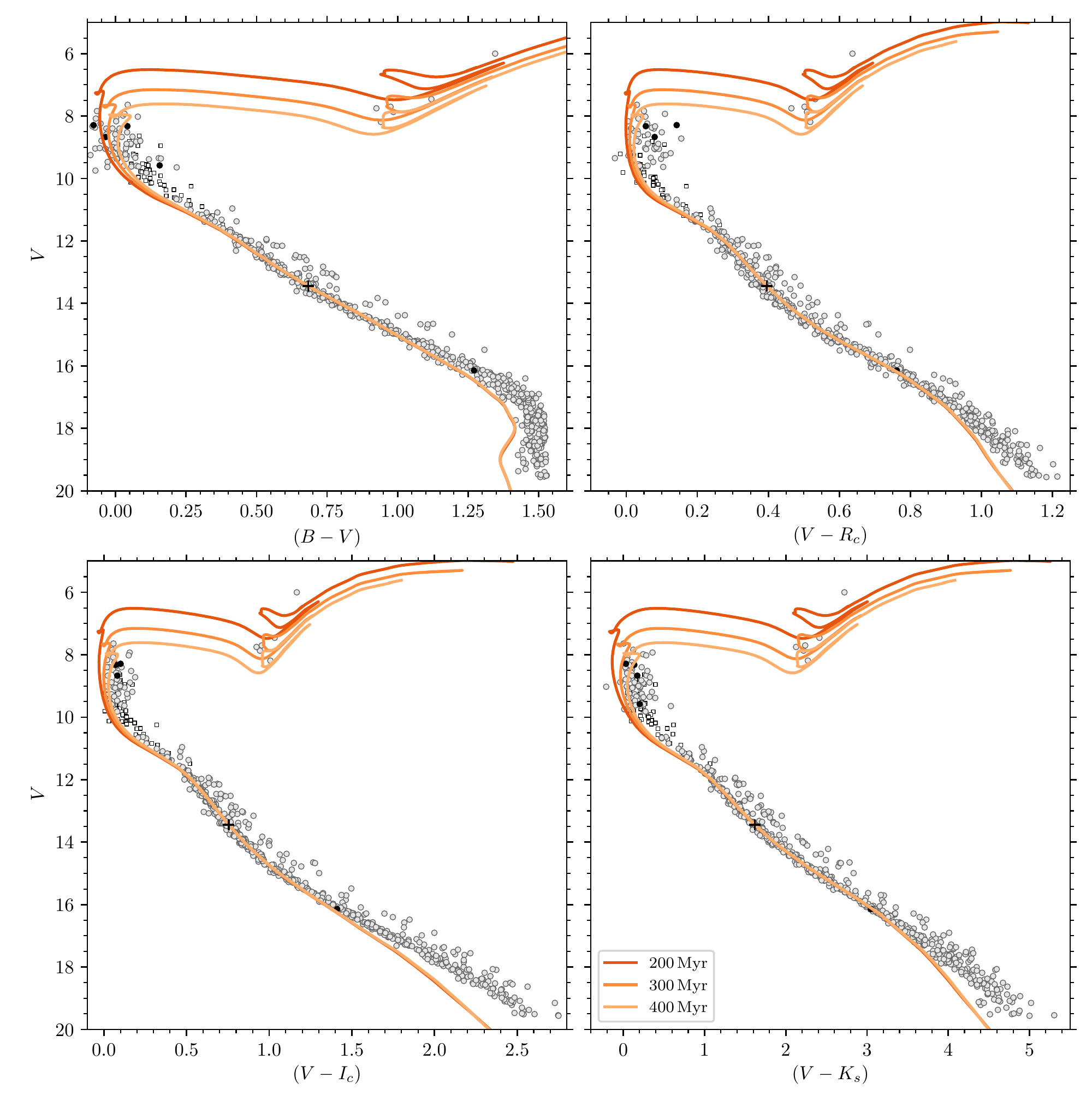}
	\caption{Same as Fig~\ref{fig:isos} but a single model of different age. The BaSTI isochrones for 200, 300, and 400\,Myr (dark to light orange) are shown and shifted by the distance modulus determined with \emph{Gaia} DR2 ($(m-M)_0=8.42$\,mag). The solar-mass model for 300\,Myr is marked with a black plus sign.}
	\label{fig:ages}
\end{figure*}

The large number of cluster members identified enables us to compare the observations in detail to theoretical isochrones. In Fig.~\ref{fig:isos} we show the cluster members from the giant branch through the turn-off down to the low-mass stars. The membership is nearly complete for the inner 1\degr region of NGC~3532 from the brightest stars in NGC~3532 down to $V=10$. In Fig.~\ref{fig:isos} a sparsely populated region of radial velocity members between $V=10$ and $V=13$ is visible. In this region the only data available come from the RAVE survey. To fill the gap we include in Fig.~\ref{fig:isos} and \ref{fig:ages}, for the inner 1\degr of NGC~3532, the members from the \emph{Gaia}/TGAS  proper motions \citep{2017A&A...601A..19G}. Due to the paucity of radial velocity members in this region it would be of interest for later studies to complete the radial velocity cluster sequence. Both the GES and the present study concentrate on the fainter regions of the CMD and have preferentially probed the cool stars.

For the comparison with isochrones in this subsection we use the already measured metallicity (Sec~\ref{sec:meh}) and distance (Sec.~\ref{sec:dist}) and will determine the age and the reddening towards NGC~3532.

We have not fitted an isochrone automatically to the data for several reasons. First, there is a well known, but not yet fully understood, deviation of the isochrones for the lowest-mass stars which is mainly an effect of the transformations between stellar model parameters and intrinsic colour used in the isochrones \citep{2015A&A...577A..42B, 2017ApJ...838..161S}. Second, we lack radial velocity members in the range $V=10$ to 11. This can be filled in with the proper motion members from \cite{2017A&A...601A..19G} but a third problem arises in the same range of the isochrones. The colour of the isochrone models differ a little from the observed colour. Finally, the aim this work is to present the cluster sequence rather than a detailed fit of the the best isochrone model.

The reddening towards NGC~3532 is known to be very small, despite its location in the Galactic disc. Previous estimates of the reddening include \cite{1980A&AS...39...11F} ($E_{B-V} = 0.042\pm0.016$\,mag), \cite{1981A&AS...43..421J} ($E_{B-V} = 0.1\pm0.04$\,mag), \cite{1988MNRAS.235.1129C} ($E_{B-V} = 0.07\pm0.02$\,mag), and C11 ($E_{B-V} = 0.028\pm0.006$\,mag). 

\begin{figure}
	\includegraphics[width=\columnwidth]{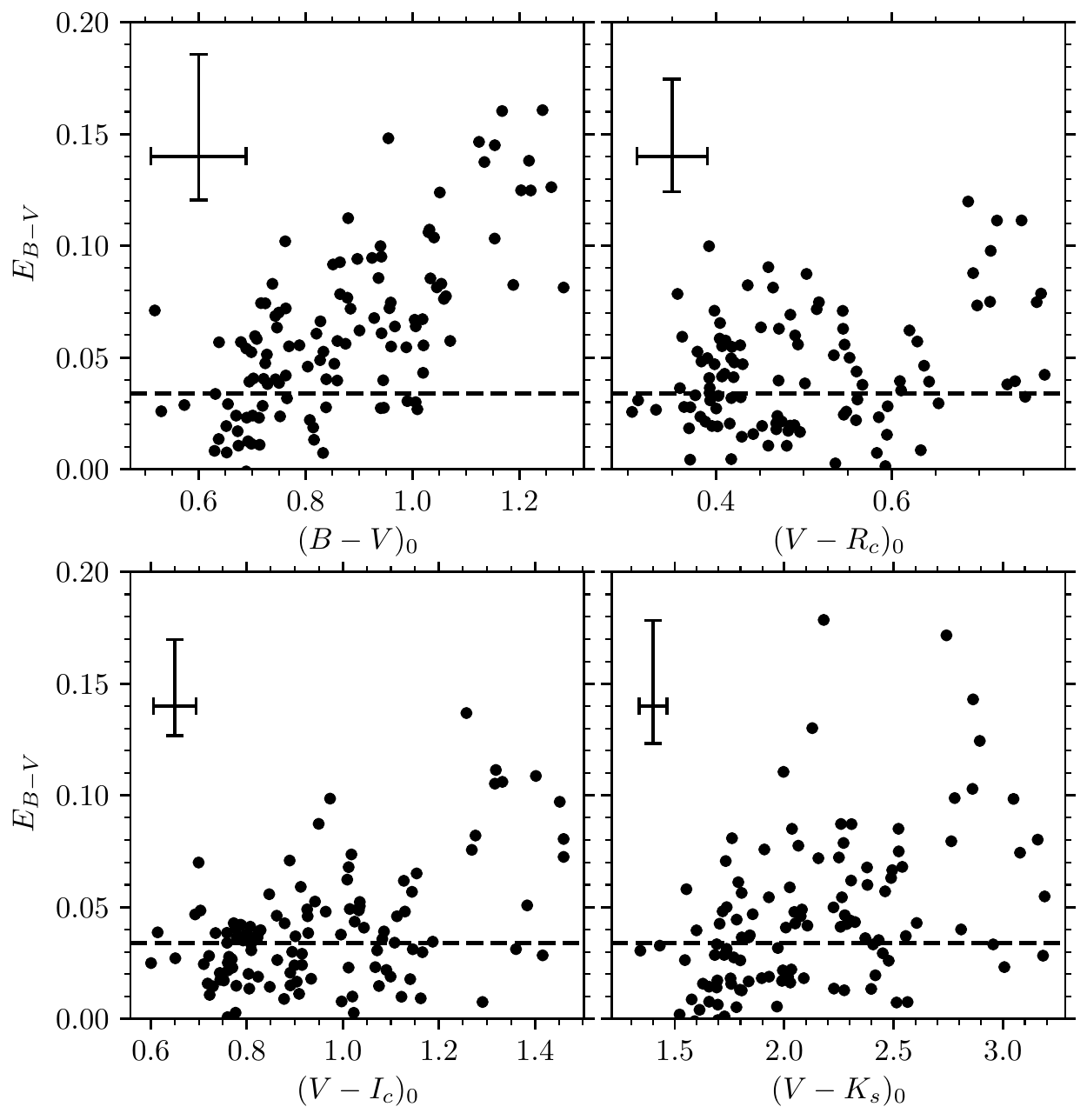}
	\caption{Reddening estimate calculated from the effective temperatures and the measured colours. For each intrinsic colour the reddening was calculated and converted to $E_{B-V}$ with the coefficients from \cite{1968nim..book..167J}. The dashed line marks the finally adopted reddening $E_{B-V}=0.034$\,mag. In the upper left of each panel we show the typical uncertainties.}
	\label{fig:reddening}
\end{figure}

In order to derive a reddening independently we used the obtained stellar parameters from the AAOmega spectra and compared them to the photometric measurements. We first calculated the intrinsic colours from the measured effective temperatures of the members of NGC~3532 by applying the $T_\mathrm{eff}$-colour relations of \cite{2005ApJ...626..465R} which are implemented in the software package \textsc{PyAstronomy}. We made use of the multi-colour photometry and calculated $(B-V)_0$, $(V-R_c)_0$, $(V-I_c)_0$, and $(V-K_s)_0$. With those values we were able to calculate the reddening $E_\mathrm{colour}$  as presented in Fig.~\ref{fig:reddening}. We used the coefficients  from \cite{1968nim..book..167J}, transformed all reddening values to $E_{B-V}$, and averaged them per colour.

Except for $(B-V)$, all other colours give a consistent extinction towards the members of NGC~3532 of $E_{B-V} = 0.034\pm0.012$\,mag. In $(B-V)$ the reddening is correlated with the intrinsic colour, a fact likely due to the colour transformations. In this colour we used only stars with $(B-V)_0<1.1$ to calculate the median reddening because those stars are less affected. The quoted reddening is the mean of the reddening values obtained from the different filter combinations, with the standard deviation as the uncertainty. The uncertainties of the determined individual colours from $T_\mathrm{eff}$ are about the same size as the calculated reddening (see Fig.~\ref{fig:reddening}) because of the average uncertainty on the effective temperature $\Delta T_\mathrm{eff}=100$\,K. This reddening estimate agrees with \cite{1980A&AS...39...11F}, \cite{1988MNRAS.235.1129C}, and C11; hence we use the determined value hereafter.

With the reddening, metallicity, and distance fixed, the only missing parameter to find the correct isochrone is the age. NGC~3532 is usually assumed to be 300\,Myr old (C11), hence we will use this value as a starting point. The known white dwarfs in NGC~3532 also constrain the age and \citep{2012MNRAS.423.2815D} estimated $300\pm25$\,Myr.
We also note that all parameters presented in Table~\ref{tab:clusterpars} are consistent with the values estimated by C11 and \cite{2012A&A...541A..41M} who found similar parameters to C11 from an isochrone fit of NGC~3532. However, \cite{2012A&A...541A..41M} did not fit either the low-mass stars or the whole giant branch.

In order to verify the isochronal age for NGC~3532 we decided to overlay multiple model isochrones with the data to find a model which best represents the data. Later we compare the data to isochrones of different ages from one particular model. For this exercise we use the low-mass isochrones from \cite{2015A&A...577A..42B} (hereafter BHAC), the BaSTI models (\citealt{2018ApJ...856..125H}), the MIST isochrones (\citealt{2016ApJ...823..102C, 2016ApJS..222....8D}), the PARSEC isochrones (\citealt{2017ApJ...835...77M}), and the YaPSI models (\citealt{2017ApJ...838..161S}). We used isochrones with [Fe/H]$=-0.1$ for all models except for BHAC, for which we had to use the solar metallicity models. All magnitudes are transformed to Johnson-Cousins and the 2MASS system \citep{2001AJ....121.2851C}\footnote{In fact we used the updated coefficients from http://www.astro.caltech.edu/$\sim$jmc/2mass/v3/transformations/.} as needed.

All of the tested models are very similar and follow the observed cluster sequence reasonably. Variations can be found on the detailed level and we note differences in order to find a model that represents the data best, moving from high to low-mass stars. The giants of NGC~3532 are represented\footnote{Except BHAC, which does not include stars of high mass.} by all models equally well, although it seems as if the YaPSI models are somewhat fainter. The same can be observed at the turn-off where the YaPSI models are $\sim0.2$\,mag fainter than the other models. Moving along the main-sequence we focus on the solar-like stars. Each CMD in Fig.~\ref{fig:isos} has the solar model (BaSTI) marked. We see that in $(B-V)$ the BaSTI model matches the position of the main-sequence somewhat better than the others under the assumption that the \emph{Gaia} distance is correct. The other models are slightly too faint. In the other three colours the isochrones differ a little from one other but seem to follow the cluster sequence well for solar-like stars. At the faint end of our observations none of the model isochrones describe the observations well. Part of the reasons for deviations from the empirical cluster sequence are the above-mentioned colour transformation issues. We suspect that the better match of the BaSTI models originates in their semi-empirical nature of their colour transformations. Regardless, we adopted the BaSTI model to discuss the age considerations because it best represents the solar-like stars under the assumption of the \emph{Gaia} distance.

\begin{table}
	\caption{Properties estimated for NGC~3532.}
	\label{tab:clusterpars}
	\begin{tabular}{lll}
		\hline
		\hline
		Property & Symbol & Value\\
		\hline
		Age & $t$ & $300\pm\sim50$\,Myr\\
		Distance modulus & $(m-M)_0$& $8.42\pm0.14$\,mag\\
		Distance & $d$ & $484^{+35}_{-30}$\,pc\\
		Reddening & $E_{(B-V)}$ & $0.034\pm0.012$\,mag \\
		Metallicity & [Fe/H] & $-0.07\pm0.1$\\
		\hline
	\end{tabular}
\end{table}

In the previous analysis we have fixed the age of NGC~3532 to 300\,Myr based on \cite{2012MNRAS.423.2815D}. To illustrate the differences between younger and older cluster models we show in Fig.~\ref{fig:ages} the BaSTI isochrones for the age of 200, 300, and 400\,Myr ([Fe/H]$=-0.1$). All isochrones are shifted with the same distance modulus and reddened the same amount. The three models differ only in the turn-off region and for giant stars. For a younger cluster the turn-off is of course bluer and at higher mass stars while for an older model the it moves to redder, lower mass stars. The 300\,Myr isochrone matches the turn-off the best. For the giants the 200\,Myr model is too bright, the 300\,Myr model describes the data well, and the 400\,Myr model is too faint. In conclusion, 300\,Myr is a good fit to the cluster sequence of NGC~3532 and we can agree that its age is 300\,Myr within a margin of, say, 50\,Myr.

\subsection{Membership count}

To infer the total number of possible cluster members we use the observed stars and their membership fraction. The distributions of photometric members observed in this work, the GES and RAVE surveys, and \emph{Gaia} DR2 were previously displayed in Fig.~\ref{fig:obs}. In the magnitude bins between $V=11$ and $V=19$ fifty per cent or more of those stars have been observed spectroscopically, with the most complete bins being those for $V=17$ and $V=18$. For $V=10$ the fraction of observed stars is lower, with only a few stars included in RAVE and \emph{Gaia}.

\begin{figure}
	\includegraphics[width=\columnwidth]{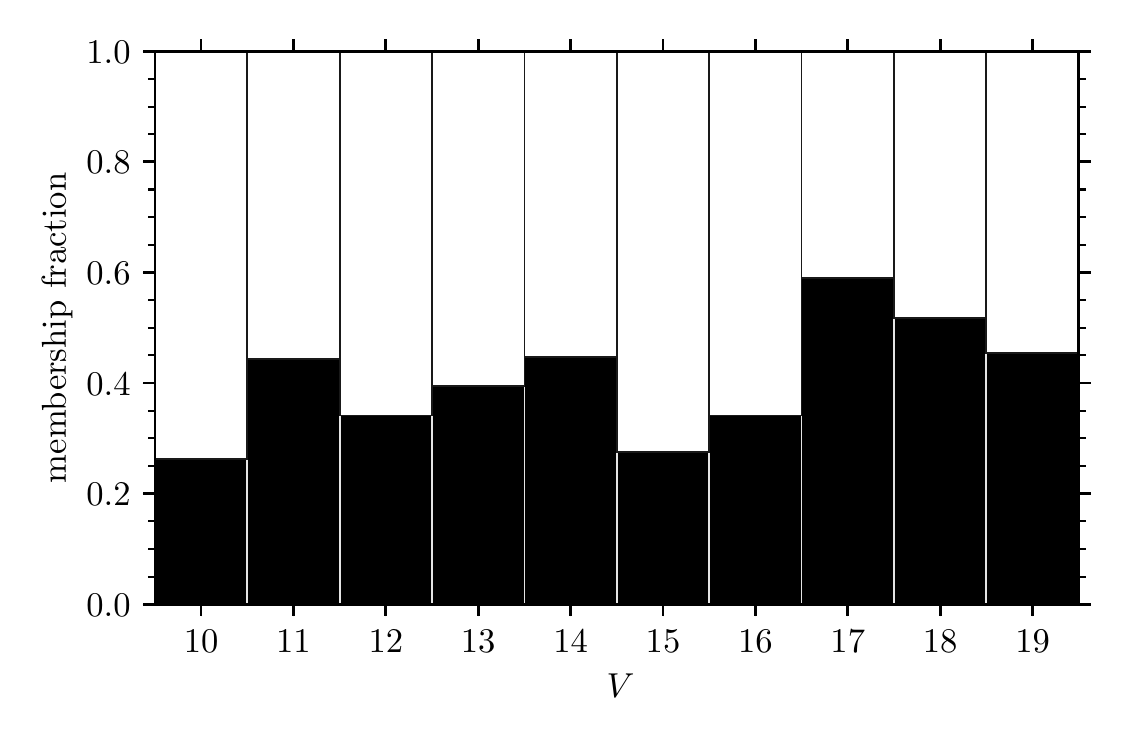}
	\caption{Membership fraction for the spectroscopically observed stars as a function of $V$ magnitude.}
	\label{fig:memberhist}
\end{figure}

Among the observed stars we find that about half are members. Although one might expect that it is easier, and therefore more likely, to find the members among the brighter stars, the fraction of non-members is nearly constant for all observed magnitude bins, as shown in Fig.~\ref{fig:memberhist}.

\label{sec:memcount}

From both the fraction of observed stars and the fraction of exclusive cluster members among those, we estimate the total number of expected cluster members. Not having complete coverage of the cluster we have had to rely on two assumptions. First, the fraction of radial velocity members among the unobserved photometric members is the same as that for the observed members. Since we selected our targets randomly from the photometric members the unobserved stars are expected to have the same distribution. Second, the membership for stars brighter than $V=10$ is complete. With the wealth of radial velocity studies at the bright end of the cluster sequence, especially the one by \cite{1981A&A....99..155G} and the recent astrometric analysis with the TGAS data \citep{2017A&A...601A..19G}, we believe that this holds true for the inner 1\degr{} of the cluster.

We extrapolate the number of members based on those assumptions. This yields an estimated number of $\sim$1000 cluster members for NGC~3532 in the studied region, an impressive number for any open cluster (c.f. \citealt{1930LicOB..14..154T}). The photometry by C11 and our radial velocity study include only the inner 1\,deg$^2$ of NGC~3532; we actually expect the cluster to reach out much further (up to 5\degr, equivalent to 15\,pc) as the first \emph{Gaia} results \citep{2017A&A...601A..19G} suggested. Many additional cluster members will likely be found outside our currently studied region. Furthermore, the \emph{Gaia} DR2 data themselves suggest 1879 cluster members within 2.31\degr  \citep{2018A&A...616A..10G}, making NGC~3532 one of the richest open clusters in that study.

\begin{figure}
	\includegraphics[width=\columnwidth]{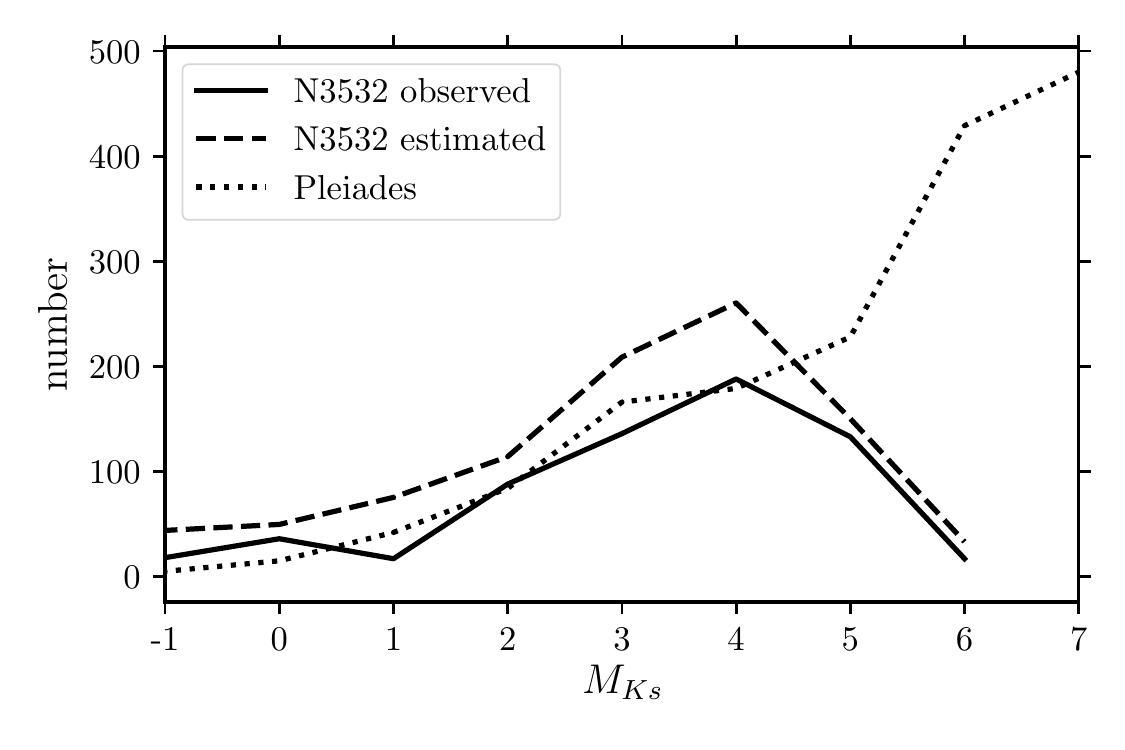}
	\caption{Luminosity functions for NGC~3532 and the Pleiades. For NGC~3532 we show the empirical luminosity function (solid line) based on our radial velocity observations and an estimate for the inner 1\degr{} of the cluster (dashed line) based on the photometric membership. The Pleiades luminosity function (dotted line) is based on \cite{2015A&A...577A.148B}.}
	\label{fig:lfunc}
\end{figure}

In order to compare the stellar content of NGC~3532 with the well-studied and very rich Pleiades cluster we constructed a luminosity function in $M_{K_s}$ for both NGC~3532 and the Pleiades. (In the latest Pleiades membership by \cite{2015A&A...577A.148B} $K_s$ is the only passband available for all stars.) We display the luminosity functions obtained in Fig.~\ref{fig:lfunc}. Three lines are displayed: first the luminosity function from the current observations of NGC~3532 (solid); second our estimate of the total NGC~3532 luminosity function based on the membership fraction of the observed stars (dashed); and third the Pleiades luminosity function (dotted).

As expected, the observed luminosity function shows features that hint at biases in the data. There are fewer stars than one would expect in the $M_{K_s}=1$ bin. This bin corresponds to the poorly observed range between $V=10$ and $V=12$, and the dip is to be expected. In the estimated luminosity function this bias is corrected, although the membership fraction is based only on a few stars. At the faint end of our observed luminosity function the number of stars drops significantly, although it is expected to rise further. Based on the comparison with the Pleiades in Fig.~\ref{fig:lfunc} we would expect $\sim 400$ members for $M_{K_s}=5$ but find only 135.

Our photometric membership information is incomplete at the faint end for two reasons. First, we included only stars common to the 2MASS survey and the photometry of C11, leading to a theoretical brightness limit of $V=19.1$, derived from the brightness limit of 2MASS. Additionally we filtered the 2MASS data to only include photometry that is good in all passbands; hence the rejection rate near the faint end is higher. Second, we used a fixed criterion to define the photometric cluster members and selected only stars up to 0.1\,mag bluer and 1\,mag brighter than the manually traced cluster sequence. As seen from Fig.~\ref{fig:cmd}, the cluster sequence in [$(B-V)$, $V$] runs nearly vertically in the CMD, binding the region of selection tightly to the manually traced cluster sequence. This may leave potential cluster members classified as photometric non-members.

Based on our estimate we conclude that NGC~3532 is likely as rich as the Pleiades, if not even richer. Seen in this context it is even more surprising that NGC~3532 was hitherto poorly studied in the astronomical literature. This open cluster is a very interesting test bed for further studies of stellar evolution especially when compared to the Pleiades. With an age of 300\,Myr it is over twice as old as the Pleiades and has only half the age of the Hyades cluster. This age range in-between the two well-studied open clusters is a critical one for studies of stellar rotation and for dynamo transitions in cool stars (e.g. \citealt{2003ApJ...586..464B}).

\section{Conclusions}
\label{sec:conclusion}

We have presented a spectroscopic study of stars in the field of NGC~3532, an open cluster embedded in the crowded field of the Galactic disc. We construct a membership list for the open cluster from our radial velocity study and \emph{Gaia} proper motions.

To select the targets for spectroscopy we constructed a photometric cluster membership list containing 2230 stars within a 1\degr{} field centred on NGC~3532. For about half of those photometric members we obtained spectra from the AAO with the fibre-fed AAOmega spectrograph and from CTIO with the Hydra-S spectrograph, and measured their radial velocities.

Combining our radial velocity measurements with data from the literature and the \emph{Gaia}-ESO and RAVE surveys we construct a radial velocity membership of NGC~3532. With the precise proper motions from \emph{Gaia} DR2 we were able to improve on that list and find 660 stars to be also proper motion members, defining a joint and exclusive membership list of these stars. Based on the fraction of observed stars and the confirmed radial velocity members we expect NGC~3532 to contain at least 1000 members within 1\degr{}, making it one of the richest clusters within 500\,pc, on par with the well-studied Pleiades. Despite the large number of cluster members presented in this work the cluster sequence is not yet complete. This may be addressed by further ground-based radial velocity observations and additional astrometry from the \emph{Gaia} mission.

We provide in Table~\ref{tab:rv} all observed radial velocities together with our computed membership probability. For use in the wider open cluster community we created a cluster sequence for NGC~3532 (Table~\ref{tab:empiso}) in various photometric colours, including the passbands of the \emph{Gaia} photometric system.

From our spectroscopic observations of NGC~3532 we find the cluster to be slightly metal-poor, with $\mathrm{[Fe/H]}=-0.07\pm0.10$. In comparison, most other studies, which focused on the giant stars, found $\mathrm{[Fe/H]}=0$ to 0.1. For a definitive statement a homogeneous analysis of giants and dwarfs with reduced uncertainties would be necessary.

In addition to abundances, we measured effective temperatures and surface gravities from the spectra. This helped to remove false-positive background giants from the sample. Furthermore, we used the effective temperature to estimate the reddening towards NGC~3532 and found $E_{B-V} = 0.034\pm0.012$\,mag, in good agreement with previously estimated values.

The precise astrometric measurements from \emph{Gaia} DR2 enabled us not only to determine an exclusive membership based on proper motions and radial velocities but also to determine the distance to NGC~3532 independently of isochrone models. Based on the parallaxes we find a distance of $484^{+35}_{-30}$\,pc ($(m-M)_0 = 8.42\pm{0.14}$\,mag).

With the metallicity, reddening, and distance known, we compared different model isochrones to the obtained cluster sequence. We showed that most modern isochrones follow the sequence well but find small differences between the models. From the BaSTI models for different ages we infer that NGC~3532 has an age of $300\pm\sim50$\,Myr.

Echoing the words of Herschel, we conclude that NGC~3532 is truly an outstanding open cluster with a very rich stellar population. 

\begin{acknowledgements}
We thank Frederico Spada for providing the interpolated YaPSI isochrones and for useful discussions.
We thank Corrado Boeche for useful discussions about \textsc{SP\_Ace}.
We thank the anonymous referee for the very helpful comments which led to improvements of the paper.
We thank D.~Zucker and C.~Lidman from AAO for performing the radial velocity observations.
Based in part on observations at Cerro Tololo Inter-American Observatory, National Optical Astronomy Observatory (2008A-0476; S.~Barnes, SMARTS consortium through Vanderbilt University; D.~James, 2008A-0512, 2008B-0248, 2010A-0281, 2010B-0492; S.~Meibom, 2011B-0322; A.~Geller), which is operated by the Association of Universities for Research in Astronomy (AURA) under a cooperative agreement with the National Science Foundation.
DJJ gratefully acknowledges support from National Science Foundation award NSF-1440254.
This research has made use of NASA's Astrophysics Data System Bibliographic Services.
This research has made use of the SIMBAD database and the VizieR catalogue access tool, operated at CDS, Strasbourg, France.
This publication makes use of data products from the Two Micron All Sky Survey, which is a joint project of the University of Massachusetts and the Infrared Processing and Analysis Center/California Institute of Technology, funded by the National Aeronautics and Space Administration and the National Science Foundation.
This publication makes use of the RAVE survey. Funding for RAVE has been provided by: the Australian Astronomical Observatory; the Leibniz-Institut für Astrophysik Potsdam (AIP); the Australian National University; the Australian Research Council; the French National Research Agency; the German Research Foundation (SPP 1177 and SFB 881); the European Research Council (ERC-StG 240271 Galactica); the Istituto Nazionale di Astrofisica at Padova; The Johns Hopkins University; the National Science Foundation of the USA (AST-0908326); the W. M. Keck foundation; the Macquarie University; the Netherlands Research School for Astronomy; the Natural Sciences and Engineering Research Council of Canada; the Slovenian Research Agency; the Swiss National Science Foundation; the Science \& Technology Facilities Council of the UK; Opticon; Strasbourg Observatory; and the Universities of Groningen, Heidelberg and Sydney. The RAVE web site is at https://www.rave-survey.org.
Based on observations made with ESO Telescopes at the Paranal Observatories under programme ID 188.B-3002 (\emph{Gaia}-Eso survey).
This work has made use of data from the European Space Agency (ESA) mission {\it Gaia} (\url{https://www.cosmos.esa.int/gaia}), processed by the {\it Gaia} Data Processing and Analysis Consortium (DPAC, \url{https://www.cosmos.esa.int/web/gaia/dpac/consortium}). Funding for the DPAC has been provided by national institutions, in particular the institutions participating in the {\it Gaia} Multilateral Agreement.
\newline
\textbf{Software:}
\textsc{PyRAF} is a product of the Space Telescope Science Institute, which is operated by AURA for NASA.
This research made use of \textsc{Astropy}, a community-developed core Python package for Astronomy \citep{2013A&A...558A..33A}.
This work has made use of the \textsc{SP\_Ace} spectral analysis tool version 1.1.
This work has made of \textsc{Topcat} \citep{2005ASPC..347...29T}.
This research made use of the following \textsc{Python} packages:
\textsc{Pandas} \citep{pandas};
\textsc{NumPy} \citep{numpy};
\textsc{MatPlotLib} \citep{Hunter:2007};
\textsc{IPython}: \citep{ipython};
\textsc{SciKit-learn}: \citep{scikitlearn};
\textsc{Seaborn}: \citep{seaborn}.
This work made use of \textsc{PyAstronomy} and \textsc{Speclite}.
\end{acknowledgements}

\bibliographystyle{aa} 

\bibliography{NGC3532-full} 

\end{document}